\documentclass[aps,amsfonts,nofootinbib,preprintnumbers,nobalancelastpage]{revtex4}
\usepackage[english]{babel}
\usepackage{amsmath,amssymb}
\usepackage{graphicx}
\usepackage[sort&compress]{natbib}
\usepackage{bbm}
\usepackage{color}
\usepackage{footnote}
\usepackage{slashed} 
\usepackage{pdflscape} 
\usepackage{multirow}
\usepackage{multirow}
\usepackage{epsfig}
\usepackage{comment}

\makeatletter
\def\endfmffile{%
  \fmfcmd{\p@rcent\space the end.^^J%
          end.^^J%
          endinput;}%
  \if@fmfio
    \immediate\closeout\@outfmf
  \fi
  \IfFileExists{\thefmffile.mp}{\immediate\write18{mpost \thefmffile}}{}
  \let\thefmffile\relax
}
\makeatother
\usepackage[normalem]{ulem}

\newcommand{\ie} {{\it i.e.}}

\newcommand {\beq} {\begin{equation}}
\newcommand {\eeq} {\end{equation}}
\newcommand {\bea} {\begin{eqnarray}}
\newcommand {\eea} {\end{eqnarray}}

\newcommand{\cf}{{\it cf. }}
\newcommand{\eg}{{\it e.g. }}

\definecolor{red1}{cmyk}{0,1,1,0.1}
\definecolor{blue1}{cmyk}{1,0,0,0}

\newcommand{\ignore}[1]{}
 \renewcommand{\bf}{\textbf}

\newcommand{\GeV}{{\rm\ GeV}}

\newcommand{\TeV}{{\rm\ TeV}}
\newcommand{\fb}{{\rm\ fb}}

\allowdisplaybreaks

\begin{document}


\title{Boosted Event Topologies from TeV Scale Light Quark Composite Partners}
\date{\today}




\author{Mihailo Backovi\'{c}} 
\email{mihailo.backovic@weizmann.ac.il} 
\affiliation{Department of Particle Physics and Astrophysics, \\ Weizmann Institute of Science, Rehovot 76100, Israel}
\affiliation{Center for Cosmology, Particle Physics and Phenomenology - CP3, Universite Catholique de Louvain, Louvain-la-neuve, Belgium}
\author{Thomas Flacke}
\email{flacke@kaist.ac.kr}
\affiliation{Department of Physics, Korea Advanced Institute of Science and Technology, \\
335 Gwahak-ro, Yuseong-gu, Daejeon 305-701, Korea} 
\author{Jeong Han Kim} 
\email{jeonghan.kim@kaist.ac.kr} 
\affiliation{Department of Physics, Korea Advanced Institute of Science and Technology, \\
335 Gwahak-ro, Yuseong-gu, Daejeon 305-701, Korea} 
\affiliation{Center for Theoretical Physics of the Universe, IBS, Daejeon, Korea}
\author{Seung J. Lee} 
\email{sjjlee@kaist.ac.kr} 
\affiliation{Department of Physics, Korea Advanced Institute of Science and Technology, \\
335 Gwahak-ro, Yuseong-gu, Daejeon 305-701, Korea} 
\affiliation{School of Physics, Korea Institute for Advanced Study, Seoul 130-722, Korea} 

%
%
%
%


\begin{abstract}
We propose a new search strategy for quark partners which decay into a boosted Higgs and a light quark.
As an example, we consider phenomenologically viable right handed up-type quark partners of mass $\sim 1 \TeV$ in composite pseudo-Nambu-Goldstone-boson Higgs models within the context of flavorful naturalness.
Our results show that $S/B > 1$ and signal significance of $\sim 7\sigma$ is achievable at  $\sqrt{s} = 14 \TeV$ LHC with 35 $\fb^{-1}$ of integrated luminosity, sufficient to claim discovery of a new particle. A combination of a multi-dimensional boosted Higgs tagging technique, kinematics of pair produced heavy objects and $b$-tagging serves to efficiently diminish the large QCD backgrounds while maintaining adequate levels of signal efficiency. We present the analysis in the context of effective field theory, such that our results can be applied to any future search for pair produced vector-like quarks with decay modes to Higgs and a light jet.
\end{abstract}


\maketitle

\section{Introduction}\label{sec:intro}
The Large Hadron Collider (LHC) has begun to explore the electroweak symmetry breaking (EWSB) scale. With a successful completion of Run I, highlighted by the discovery of the Higgs boson~\cite{Aad:2012tfa,Chatrchyan:2012ufa}, the Standard Model (SM) is now complete. 
The Higgs boson accounts for the EWSB, generates masses of fermions, provides an explanation for the short range of the weak force, as well as unitarizes the $W$-boson scattering cross section. However, within the SM there is no explanation for why the Higgs boson mass itself is $O(100 \GeV)$. The naive expectation from perturbation theory shows that the Higgs mass should be close to the ultra-violet (UV) scale of the theory, due to the large couplings of the Higgs to the top quark ($i.e.$ the hierarchy problem). There is a-priori no physical principle which prevents the Higgs mass from being finely tuned, although it is extremely uncommon to encounter such finely tuned quantities in nature. The latter prompted much of the theoretical work in the past decades to seek the explanation for the hierarchy problem within the scope of the ``naturalness'' paradigm. 

There are two common ``natural'' solutions to the hierarchy problem. The first is to introduce additional symmetries to protect the Higgs mass from large corrections. The second is to model the Higgs boson as a composite object \cite{Kaplan:1983fs,Kaplan:1983sm,Georgi:1984ef,Banks:1984gj,Georgi:1984af,Dugan:1984hq,ArkaniHamed:2002qy,Contino:2003ve,Giudice:2007fh,Barbieri:2007bh,Panico:2011pw,DeCurtis:2011yx,Marzocca:2012zn,Bellazzini:2012tv,Panico:2012uw,Bellazzini:2014yua}, such that the Higgs mass becomes irrelevant above some dynamically generated compositeness scale, analogous to the pion mass in Quantum Chromo Dynamics (QCD).
From the low energy effective theory point of view, both mechanisms  introduce additional degrees of freedom ($i.e.$ top partners) to the SM\footnote{For solutions within composite Higgs models which do not require top partners \cf Refs.~\cite{Galloway:2010bp,Cacciapaglia:2014uja}.}, which cancel the top loop induced quadratic divergences in the Higgs mass. The top partners can be scalars, as in the case of supersymmetry, or fermions,  as in the case of composite Higgs models.  Together, the two mechanisms provide a ``litmus test'' for the naturalness paradigm.

The LHC is finally able to put naturalness to a meaningful test, where most of the experimental effort has been focused on searches for top partners \cite{Chatrchyan:2013uxa, ATLAS-CONF-2014-036}. The fact that no super-partners have been observed at the LHC is already pushing the supersymmetric models into a 
tuned regime.  However, as the bounds on the scalar top partner mass increase, there have been several attempts to relax the bounds on the top partners via compressed/stealth spectrum, R-parity violation,  Dirac gauginos, split families, etc.~\cite{Fan:2011yu,Hall:2011aa,Papucci:2011wy,LeCompte:2011fh,LeCompte:2011cn,Csaki:2011ge,Fan:2012jf,Kribs:2012gx,Craig:2012di,Evans:2012bf,Dreiner:2012gx,Dreiner:2012sh,Mahbubani:2012qq,Blanke:2013uia,Han:2013lna,Kribs:2013oda}. Composite Higgs models are in a similar situation, although the bounds on the spin $1/2$ partners in such models are somewhat milder compared to the already existing bounds from LEP and Tevatron constraints on the oblique parameters ~\cite{Grojean:2013qca,Ciuchini:2013pca}. With the increased center of mass energy, Run II of the LHC will soon be able to cover the interesting region of parameter space of composite top partners~\cite{Backovic:2014uma}.

An interesting avenue to bypass existing bounds is to employ non-trivial flavor structure for top partners \footnote{ Commonly referred to as
``flavorful naturalness''~\cite{Blanke:2013uia}.}, where a large mixing is allowed between the right-handed (RH) top and RH charm partners.
The basic idea comes from a simple observation that scalar top partners ($i.e.$ stops) need not be mass eigenstates in order to cancel the large SM loop corrections to the Higgs mass. Instead, a stop flavor eigenstate made up of a stop-like and scharm-like mass eigenstates can serve the same purpose~\cite{Mahbubani:2012qq,Blanke:2013uia}. 
An analogous approach has recently been applied to composite Higgs models for light non-degenerate composite quarks~\cite{Delaunay:2013pwa}.
The analysis focused on the Minimal Composite Higgs model (MCHM)~\cite{Agashe:2004rs} based on the coset structure $SO(5)$$/$$SO(4)$, in which the Higgs doublet was realized as a pseudo-Goldstone boson.

Implementing non-degenerate composite quarks into composite Higgs models without conflict with the existing bounds from flavor physics and electro-weak (EW) precision observables is a non-trivial task. 
However, Ref.~\cite{Galon:2013jba} showed that flavor alignment allows models with non-degenerate light generation partners to satisfy the constrains from flavor physics observables \footnote{As shown in the case of original supersymmetric flavorful naturalness, mixing between left-handed partners of top and charm give rises to more severe constraints from FCNC processes, and it was preferred to choose the mixing through the RH partners for the simplicity. The situation is similar for  composite Higgs models. Thus, we focus on the RH up-type partners in our analysis.}. In addition, models with custodial parity~\cite{Agashe:2006at,Contino:2006qr} have been shown to be consistent with the constraints from EW precision tests~\cite{Delaunay:2010dw, Redi:2011zi}. Collider implications for such scenario have also been studied in Refs.~\cite{Delaunay:2011vv, Redi:2013eaa}.

Ref.~\cite{Delaunay:2013pwa} studied the implications of non-degenerate composite partners of the first two generation quarks for LHC phenomenology and derived the LHC bounds on fermionic resonances in the $SO(4)$ fourplet representations.  In particular, Ref.~\cite{Delaunay:2013pwa} showed that, without assuming degenerate compositeness parameters, the fourplet RH up-quark partners have to be  heavier than $\sim2$ TeV or the degree of compositeness of RH up quark has to be very small. In the latter case,  a lower mass bound of $\sim 530$~GeV still applies. At the same time,  the fourplet RH charm quark component can be mostly composite and its partners can be as light as $600$ GeV even with a large degree of right-handed compositeness.

Contrary to fourplet partners, $SO(4)$ singlet partners are barely constrained by the LHC Run I searches. Ref. ~\cite{Flacke:2013fya} recently obtained the first non-trivial bound on $SO(4)$ singlet partners utilizing the $h\rightarrow \gamma\gamma$ results from ATLAS~\cite{TheATLAScollaboration:2013eia}. However, the bound ($i.e.$ the RH up-type partner mass $M_{U_h} > 310$ GeV) is very mild as the experimental searches were not designed to search for Higgs bosons arising from composite light quark partner decays.

The main focus of this paper is to design a dedicated search for singlet partners of light quarks, and study the potential of such searches to discover the quark partners at the Run II of the LHC. For the purpose of illustration, we study right-handed up-type quark partners, which are QCD pair-produced and decay dominantly into a Higgs boson and an up-type quark. We design the analysis in an effective theory framework, such that -- although being motivated by composite quark partner searches -- our results can be applied to any heavy vector-like quark model in which the vector-like quark has a decay channel into a Higgs and a light quark.

We focus on the potential of LHC Run II to probe light quark partners of mass $\sim 1 \TeV$, where the decays of light quark partners typically result in boosted Higgs bosons. In order to increase the signal rate, we consider only the decays of the Higgs boson to a $b\bar{b}$ pair.  Seemingly complicated, such final states are particularly interesting, as traditional event reconstruction techniques fail.  Due to the large degree of collimation of Higgs decay products, methods of Higgs tagging via ``jet substructure''  need to be employed \cite{Butterworth:2008iy}. In addition, the boosted di-Higgs event topology accompanied by two light jets offers a myriad of handles on large SM backgrounds. As we will show in the following sections, a combination of kinematic constraints of pair produced heavy particles, boosted Higgs tagging and double $b$-tagging is able to achieve a signal to background ratio $S/B > 1$ for light quark partner masses of 1~TeV. The same analysis shows that signal significance of $\sim 7 \sigma$ can be achieved with $35 \fb^{-1}$ of integrated luminosity, sufficient to claim a discovery. 

 For the purpose of boosted Higgs tagging,  we use the Template Overlap Method (TOM) ~\cite{Almeida:2011aa,Almeida:2010pa, Backovic:2012jj, Backovic:2013bga}. We propose a new form of overlap analysis which utilizes both Higgs template tagging and top template tagging in order to optimize the rejection of SM backgrounds while maintaining sufficient signal efficiency. The ``multi-dimensional'' TOM tagger compares the likelihood that a boosted jet is a Higgs to the likelihood that a boosted jet is a top quark, whereby a Higgs tag assumes that a jet is sufficiently Higgs like and not top like. Furthermore, we find that requiring at least one $b$-tag in each of the Higgs tagged jets significantly improves signal purity, especially with respect to large multi-jet backgrounds.  
 
We organized the paper in three sections. Sec.~\ref{sec:model} summarizes the theoretical framework of MCHM with partially composite RH up-type quark partners and introduces the effective model of the light up-type quark partners. In Sec.~\ref{sec:model} we also discuss the diagonalization of mass matrices, calculation of the couplings in the mass eigenbasis and other relevant parameters which enter the effective parametrization used throughout the paper.
Sec.~\ref{sec:results} deals with a phenomenological study of LHC Run II searches for up-type quark partners. We propose and discuss in detail a set of observables which can be used to efficiently detect and measure the partners at $1 \TeV$ mass scales, as well as present results on $S/B$ and signal significance using our cutflow proposal. We conclude in Sec.~\ref{sec:conclusions}. A brief discussion of models in which the quark partner is not dominantly RH can be found in the Appendix.

\section{Partially composite light quark partners}\label{sec:model}

In this article we focus on the MCHM based on the coset structure $SO(5)$$/$$SO(4)$.  We follow the conventions and notation of Ref.~\cite{Delaunay:2013pwa} based on the Callan-Coleman-Wess-Zumino (CCWZ) formalism \cite{ccwz1, ccwz}. The Higgs multiplet is non-linearly realized as the Goldstone Boson multiplet of the $SO(5)\times U(1)_X\rightarrow SO(4)\times U(1)_X\sim SU(2)_L\times SU(2)_R\times U(1)_X$ breaking. Gauging the $SU(2)_L$ and $Y=T^3_R+X$ assigns the correct $SU(2)\times U(1)_Y$ quantum numbers to the Higgs multiplet, which is parameterized by the Goldstone boson matrix. In unitary gauge, it reads \cite{Delaunay:2013pwa, DeSimone:2012fs}
\beq
U_{\rm gs}=\left(\begin{array}{ccccc}
                1 & 0 & 0 & 0 & 0 \\
		 0 & 1 & 0 & 0 & 0 \\
		 0 & 0 & 1 & 0 & 0 \\
		 0 & 0 & 0 & \cos{\frac{h+\langle h\rangle}{f}} & \sin{\frac{h+\langle h\rangle}{f}} \\
		 0 & 0 & 0 & -\sin{\frac{h+\langle h\rangle}{f}} & \cos{\frac{h+\langle h\rangle}{f}}
               \end{array}\right)\,,
               \label{gmatrU}
\eeq
where $\langle h \rangle$ is vacuum expectation value of the non-linearly realized Higgs field which is related to the Standard Model vacuum expectation value by $246 \GeV \equiv v = f \sin(\langle h \rangle / f) \equiv  f \sin(\epsilon)$.

In composite Higgs models, the Higgs transforms non-linearly under the global spontaneously broken symmetry group, while elementary fermions transform linearly. Yukawa-type interactions of purely elementary quarks (and leptons) with the Higgs are hence forbidden. However, the strongly coupled sector is expected to contain QCD charged fermionic resonances ($i.e.$ ``quark partners'') at or below a scale $\Lambda \sim 4 \pi f$ which can have Yukawa-type couplings with elementary quarks and the Goldstone boson matrix (which contains the Higgs). Electroweak symmetry breaking then yields mass mixing terms between the composite quark partners and the elementary quarks such that the lightest quark mass eigenstates (which are identified with the SM-like quarks) are partially composite. The mass spectrum and couplings of the SM-like quarks and their heavy partners to electroweak gauge bosons and the Higgs depend on the $SO(5)$ representations in which the elementary quarks and the heavy partner quarks are embedded. 
For concreteness, here we focus on one minimal embedding.

The elementary left-handed and right-handed quarks are embedded into incomplete $\bold 5$ representations of $SO(5)$ 
\bea
\bar{q}^U_L=\frac{1}{\sqrt{2}}\left(-i\bar{d}_L\, , \bar{d}_L\, , -i\bar{u}_L\, , -\bar{u}_L\, , 0\right) \,\,\,\,\, & , & \,\,\,\,\,\, \bar{q}^D_L=\frac{1}{\sqrt{2}}\left(i\bar{u}_L\, , \bar{u}_L\, , -i\bar{d}_L\, , \bar{d}_L\, , 0\right),\\
\bar{U}_R^5=\left(0,0,0,0,\bar{u}_R\right) \,\,\,\,\, & , & \,\,\,\,\,\, \bar{D}_R^5=\left(0,0,0,0,\bar{d}_R\right),
\eea
with a $U(1)_X$ charge of $2/3$ for $q^U_L$ and  $-1/3$ for $q^D_L$.
The lightest composite quark partner resonances are assumed to be in the $\bold 5$ of $SO(5)$ as well
\beq
     \psi^U=\left(\begin{array}{c}
                Q^U\\
		  \tilde{U}
               \end{array}\right)
               =\frac{1}{\sqrt{2}}\left(\begin{array}{c}
                iD^u-iX_{5/3}\\
		 D^u+X_{5/3}\\
		 iU^u+iX_{2/3}\\
		 -U^u+X_{2/3}\\
		 \sqrt{2}\tilde{U}
               \end{array}\right) \,, \,\,\,\,\,\,\,\,\,\, 
      \psi^D=\left(\begin{array}{c}
                Q^D\\
		  \tilde{D}
               \end{array}\right)
               =\frac{1}{\sqrt{2}}\left(\begin{array}{c}
                -iU^d+iX_{-4/3}\\
		 U^d+X_{-4/3}\\
		 iD^d+iX_{-1/3}\\
		 -D^d+X_{-1/3}\\
		 \sqrt{2}\tilde{D}
               \end{array}\right),               
\eeq 
with $U(1)_X$ charge of $2/3$ for $\psi^U$ and  $-1/3$ for $\psi^D$.

Using the CCWZ prescription we can construct the fermion Lagrangian of the model which reads
\beq
\mathcal{L}=\mathcal{L}_{\rm comp}+\mathcal{L}_{\rm el, mix}\,,
\label{eq:defgenmod}
\eeq
with
\bea
\mathcal{L}_{\rm comp}&=&\left(i\ \bar{Q}^U(D_\mu +i e_\mu) \gamma^\mu Q^U + i \bar{\tilde U}\slashed{D}\tilde U
 -M^U_4\bar{Q}^UQ^U -M^U_1\bar{\tilde{U}}\tilde{U}\right.\nonumber\\
 &&\left.
 +\left(i c_{L,R}^U \bar{Q}^{U\,\,i}_{L,R} \gamma^\mu d^i_\mu \tilde{U}_{L,R}+\mbox{h.c.}\right)\right) + (U\rightarrow D)\, ,
\eea
where  $e_\mu$ and $d^i_\mu$ are the CCWZ connections (\cf Appendix~A of Ref.~\cite{Delaunay:2013pwa} for the explicit expressions), $M^{U,D}_{1,4}$ and $c^{U,D}_{L,R}$ are matrices in flavor space, and 
\beq
 \mathcal{L}_{\rm el,mix}=i\ \bar{q}_L\,\slashed{D}q_L+i\ \bar{u}_R\,\slashed{D}u_R+i \bar{d}_R\,\slashed{D}d_R+\left(-y^U_L f\bar{q}^U_L U_{\rm gs} \psi^U_R
 -y^U_R f  \bar{U}^5_R U_{\rm gs} \psi^U_L+\mbox{h.c.}\right)+\left(U\rightarrow D\right),\label{Lelmixpc}
\eeq
where the pre-Yukawa couplings $y^{U,D}_{L,R}$ are matrices in flavor space. 

Typically, the composite sector is assumed to be flavor-blind in order to avoid constraints from flavor changing neutral currents (\cf \, \eg Ref.~\cite{Redi:2011zi}). In such a setup, the flavor structure only enters via the pre-Yukawa couplings, and the partners of the different SM quark flavors are mass degenerate, up to Yukawa-suppressed corrections. However, as has been pointed out in Ref. \cite{Gedalia:2012pi}, partners are allowed to be non-degenerate within models of flavor alignment~\cite{Fitzpatrick:2007sa, Csaki:2009wc}. In this article we allow for non-degenerate quark partner masses $M^{U,D}_{1,4}$ and treat them as free parameters.

LHC run I established various constraints on the different quark partners already:\footnote{All bounds quoted refer to QCD pair production and subsequent decay of the quark partners. This production channel only depends on the mass of the quark partner and is therefore rather model-independent. The various partners can also be single-produced via electro-weak interactions. The mass bounds from such channels can be more stringent in some part of the parameter space (\cf \eg \cite{Delaunay:2013pwa, DeSimone:2012fs}) but the production cross section for these processes depends on the model parameters $y^{U,D}_{L,R}, c^{U,D}_{L,R}$ such that these constraints can be alleviated.}
\begin{itemize}
\item The top partner multiplet $Q^U_3$ contains a charge $5/3$ particle $X^T_{5/3}$ as the lightest member with a mass $M_4$. Its decay channel  $X^T_{5/3}\rightarrow W^+ t$ yields a same-sign dilepton signal which has not been observed, yet. This results in a lower mass bound of $\left(M^U_{4}\right)_3 > 800 \GeV$ established by CMS \cite{Chatrchyan:2013wfa}.
\item The singlet top partner $\tilde{T}\equiv \tilde{U}_3$ (as well as the the charge 2/3 partners in $Q^U_3$ multiplet) has decay channels into $tZ$, $th$, and $Wb$. CMS established a lower bound on the mass of a charge 2/3 partner of 687~-~782~GeV \cite{Chatrchyan:2013uxa}, with the strongest bound applying if $\tilde{T}\rightarrow tZ$ is the dominating decay. The analogous ATLAS bounds are $\sim$~350~-~810~GeV \cite{Aad:2014efa}.
\item 3rd family charge -1/3 partners can decay into $bZ$, $bh$, and $Wt$. CMS constrained their mass to lie above  582~-~785~GeV, again depending on the branching ratios \cite{CMS:2013una,CMS:2012hfa}.\footnote{Again, the bounds are strongest when the branching ratio into $Zb$ is large. However, a recent CMS study \cite{CMS:2014bfa} focussed on the the all-hadronic channel $pp\rightarrow B\bar{B}\rightarrow hbh\bar{b}\rightarrow b\bar{b} b b \bar{b}\bar{b}$ and showed that limits are improved when making use of jet-substructure techniques. Assuming 100 \% branching ratio of $B\rightarrow hb$, \cite{CMS:2014bfa} obtained a lower bound on the mass of  846~GeV.} The current ATLAS  lower mass bound on the charge $-1/3$ partners is $\sim$~350~-~800~GeV \cite{Aad:2014efa}.
\item Bounds on partners in the multiplets $Q^U_{1,2}$  have been studied in detail in Ref.~\cite{Delaunay:2013pwa}, where a bound of $\left(M^U_{4}\right)_{1,2} > 530 \GeV$ for QCD pair produced partners was established, which also applies to partners in the $Q^D_{1,2}$ multiplets. These bounds on light quark partners are weaker than the bounds on 3rd generation quark partners. Third generation partners decay into electroweak gauge bosons (or a Higgs) and a third generation quark, leading to final states which can be efficiently ``tagged'' at the LHC and hence allow to reduce or eliminate the numerous SM backgrounds. On the other hand, partners of light quarks decay into light quark flavors which are significantly more difficult to distinguish from the SM background channels.
\item So far, the most unconstrained partners are the light quark singlet partners $\tilde{U}_{1,2}$ and $\tilde{D}_{1,2}$. The dominant decay mode into $ h j$,  leads to a (potentially large) di-Higgs signature which has not been searched for at LHC run I.\footnote{ATLAS \cite{Aad:2014xxy,ATLAS:2014xxx} and CMS \cite{CMS:2014eub,CMS:2013eua}  published results on di-Higgs signals which result from the decay of a heavy resonance (KK-graviton or, respectively, a heavy Higgs), but these searches do not apply to the di-Higgs signal considered here, as the sum of the invariant mass of the decay products does not form a resonance in our case.} The only constraint we are aware of has been obtained in Ref.~\cite{Flacke:2013fya}, where the absence of $h\rightarrow \gamma\gamma$ decays with  high $p^{\gamma\gamma}_T$  has been used to establish a bound of $M_1 > 310$~GeV.
\end{itemize}

In this article, we study the discovery reach for the weakest constrained and therefore potentially lightest quark partner at LHC run II: a light-quark $SO(4)$ singlet partner. Focussing on the singlet partner, the model defined in Eq.~(\ref{eq:defgenmod}) can be simplified. For simplicity, we take the  limit $M_4 \gg M_1$, and discuss the model for the up-partner only. Note that the phenomenology of $d,s,c$ partners is analogous.\footnote{In this article we focus on parameter independent bounds which arise from QCD pair production of quark partners. For (parameter dependent) single production, the quark flavor affects the production cross section (\cf \cite{Flacke:2013fya}).}

Under these simplifying assumptions, the Lagrangian of the up-quark sector following from Eq.~(\ref{eq:defgenmod}) is \cite{Flacke:2013fya}
\beq
\begin{split}
\mathcal{L}= &\,\, i \bar{\tilde U}\slashed{D}\tilde U -M_1\bar{\tilde{U}}\tilde{U} +  i\ \bar{q}_L\,\slashed{D}q_L +i\ \bar{u}_R\,\slashed{D}u_R\\
&
-\left[-\frac{y_L}{\sqrt{2}} f \bar{u}_L  \sin\left(\frac{h+\langle h\rangle}{f}\right) \tilde{U}_R
 + y_R f\bar{\tilde{U}}_L   \cos\left(\frac{h+\langle h\rangle}{f}\right)  u_R+\mbox{h.c.}\right].\label{pcLag1}\
\end{split}
\eeq 
Expanding around the vacuum expectation value $ \langle h \rangle$ yields the effective quark mass terms
\beq
\mathcal{L}_m=-(\bar{u}_L, \bar{\tilde{U}}_L) M_u \left(\begin{array}{c} u_R\\ \tilde{U}_R \end{array}\right)+ \mbox{h.c.}
\mbox{\hspace{8pt} with \hspace{8pt}}
M_u= \left(
\begin{array}{cc}
0& -\frac{y_L}{\sqrt{2}}f \sin \epsilon\\
y_R f \cos\epsilon & M_1
\end{array}
\right)
\equiv \left(
\begin{array}{cc}
0& m_L\\
m_R & M_1
\end{array}
\right).
\eeq
Note that the effective mass terms $m_L$ and $m_R$ arise from the left- and right-handed pre-Yukawa mass terms which have inherently different symmetry properties. The $y_L$ coupling links a fundamental fourplet to a composite $SO(4)$ singlet while the $y_R$ coupling links a fundamental singlet to a composite fourplet. Therefore, $y_L$ and $y_R$ are independent parameters which are not required to be of the same order of magnitude by naturalness. For simplicity, we choose $y_R \gg y_L$ here, and discuss consequences of the opposite limit $y_R \lesssim y_L$ in Appendix \ref{app:yLlyR}. 

For $y_R \geq y_L$, the mixing mass terms have a hierarchy $m_R \gg m_L$. The eigenvalues of the squared mass matrix are
\bea
M^2_{u_l}&=&\frac{m_L^2m_R^2}{M^2_1+m_L^2+m_R^2}\left[1+\mathcal{O}\left(\frac{m_L^2 m_R^2}{\left(M_1^2+m_L^2+m_R^2\right)^2}\right)\right]\approx \frac{m_L^2m_R^2}{M^2_{U_h}}\label{Mul}\, ,\\
M^2_{U_h}&=&\left(M_1^2+m_L^2+m_R^2\right)\left[1+\mathcal{O}\left(\frac{m_L^2 m_R^2}{\left(M_1^2+m_L^2+m_R^2\right)^2}\right)\right]\approx \left(M_1^2+m_R^2\right),
\eea
where the lighter eigenvalue $M_{u_l}$ is to be identified with $m_u$, implying $|m_Lm_R|/M_1^2\ll 1$. The bi-unitary transformation which diagonalizes the mass matrix is  a rotation by $\varphi_{L,R}$ on the left- and right-handed up-quarks where
\beq
\tan{\varphi_{R}}\approx \frac{m_{R}}{M_1}\gg \tan{\varphi_{L}}\approx \frac{m_{L}}{M_1}\,.
\label{eqmixangles}
\eeq

The couplings of the mass eigenstates to the $Z$ bosons follow from rewriting 
\beq
\mathcal{L}_Z =  (\bar{u}_L , \bar{\tilde{U}}_L) 
\left[
\frac{g}{2 c_w}\left(
\begin{array}{cc}
1 & 0 \\
0 & 0
\end{array}
\right)-\frac{2 g}{3}\frac{s^2_w}{c_w}\cdot \mathbbm{1}\right] \slashed{Z}
\left(\begin{array}{c}
u_L\\
\tilde{U}_L
\end{array}
\right) - \frac{2 g}{3}\frac{s^2_w}{c_w}(\bar{u}_R, \bar{\tilde{U}}_R)\slashed{Z}\cdot \mathbbm{1}\left(\begin{array}{c}
u_R\\
\tilde{U}_R
\end{array}
\right), 
\label{ZLag}
\eeq
in the mass eigenbasis  $(u_l,U_h)$. Note that the couplings arising from the $U(1)_X$ gauge couplings are universal. A rotation into the mass eigenbasis of these terms does not induce any ``mixed'' interactions of the $Z$ to $u_l$ and $U_h$ and leaves the $Z$ couplings to right-handed light quarks unaltered.
Mixing in the left-handed sector induces non-universality of the light quark couplings to the $Z$, but the correction to the left-handed coupling is of order $\sin^2\varphi_L \sim m^2_L/M^2_1 \ll m_u/M_1\sim \mathcal{O}(10^{-6})$, such that corrections to the hadronic width of the $Z$ are negligible \footnote{For $d,s,c$ partners, the analogous corrections are $\ll$  $10^{-6}, 10^{-4}, 10^{-3}$ such that no bounds apply as long as $y_R\geq y_L$.}. The ``mixed'' coupling of the $Z$ to $u_l$ and $U_h$ in the left-handed sector is
\beq
g^{L}_{U_h u_l Z} = g \frac{\cos \varphi_L\sin \varphi_L}{2 c_w}\approx  \frac{g}{2 c_w}\frac{m_L}{M_1}\,.\label{Zeff}
\eeq
Analogous to the neutral current, the mass mixing in the left-handed sector also induces negligible corrections to the $Wud$ vertex and a ``mixed'' coupling between the $W$, $U_h$, and $d$:
\beq
g^{L}_{U_h d_l W} = \frac{g}{\sqrt{2}} \sin \varphi_L\approx  \frac{g}{\sqrt{2}}\frac{m_L}{M_1}\,.\label{Weff} 
\eeq

The Higgs couplings to the quark mass eigenstates follow from expanding Eq.~(\ref{pcLag1}) to first order in $\epsilon$ and subsequent rotation into the mass eigenbasis. In the gauge eigenbasis the Yukawa terms read 
\beq
\mathcal{L}_{\rm Yuk} =  -\frac{\lambda_L}{\sqrt{2}} h \bar{\tilde{U}}_R u_L-\frac{\lambda_R}{\sqrt{2}} h \bar{\tilde{U}}_L u_R+\mbox{h.c.} \,,
\eeq
with 
\beq
\lambda_L = -y_L \cos(\epsilon) \,\,\,\, \lambda_R = - \sqrt{2} y_R \sin(\epsilon)\,.
\eeq
Rotating into the mass eigenbasis, the mixing Yukawa interactions
\beq
\mathcal{L}_{\rm Yuk,mix} =  -\frac{\lambda^{\rm mix}_L}{\sqrt{2}} h \bar{U}_{h,R} u_{l,L}-\frac{\lambda^{\rm mix}_R}{\sqrt{2}} h \bar{U}_{h,L} u_{l,R}+\mbox{h.c.} \,,
\eeq
are 
\beq
\lambda^{mix}_L = -y_L \cos(\epsilon) \cos\varphi_L \cos\varphi_R \,\, , \,\, \lambda^{mix}_R = - \sqrt{2} y_R \sin(\epsilon) \cos\varphi_L \cos\varphi_R\,.
\label{lambdapceff}
\eeq
In the regime $y_L\ll y_R$ considered here, the mixing couplings to $h,W,Z$ which are proportional to $y_L$ can be neglected, and the model is described by the simple effective action 
\beq
\mathcal{L}_{\rm eff}=\mathcal{L}_{\rm SM}+\bar{U}_h\left(i\slashed{\partial}+ e \frac{2}{3}\slashed{A} -g\frac{2}{3}\frac{s^2_w}{c_w}\slashed{Z}+ g_3 \slashed{G}  \right)U_h - M_{U_h}\bar{U}_hU_h
-\left[\frac{\lambda^{mix}_{R}}{\sqrt{2}}h\bar{U}_{h,L}u_{l,R}+\mbox{h.c.}\right]\,. \label{Lpceff}
\eeq
The Lagrangian in Eq.~\eqref{Lpceff} and the definition of the effective coupling of Eq.~\eqref{lambdapceff} is valid for up-type quark partners. The analogous calculation for down-type partners yields the same Lagrangian with the charge factors $2/3$ being replaced by $-1/3$ as directly follows from the $U(1)_X$ charge assignments.

\bigskip

The phenomenology of this model is particularly simple:
\begin{itemize}
\item The partner state $U_h$ carries color charge and can therefore be produced via QCD pair production.\footnote{For a large value of $\lambda^{\rm mix}_R\gtrsim g_s$ and depending on the partner quark flavor, additional production channels exist which have been discussed in Ref.\cite{Flacke:2013fya}, however here, we focus on the parameter independent QCD pair production.}
\item The dominant decay channel for the quark partner is $U_h\rightarrow u h$.\footnote{Decays into $Z u$ and $W d$ are suppressed in the regime $y_L \ll y_R$ which is described by the effective Lagrangian Eq.~(\ref{Lpceff}). The decays are only present in the regime $y_L \gtrsim y_R$ with branching ratios  $\Gamma_{U_h\rightarrow h u} : \Gamma_{U_h\rightarrow Z u} : \Gamma_{U_h\rightarrow W d}$ of $1:1:2$ in the limit $y_L \gg y_R$. For a detailed discussion \cf Appendix~\ref{app:yLlyR}}
\end{itemize}
This model hence predicts $p p \rightarrow U_h \bar{U}_h \rightarrow h h j j$ as a distinct signature at the LHC. In the following sections, we will explore the prospects for discovery of such signals at the LHC Run II, with the focus on partner masses of $\sim 1 \TeV$. 

In our model, the dominant branching ratio to $U_h \rightarrow uh$ is a consequence of the fact that the quark partner is an $SU(2)$ singlet, where we assumed $y_R \gg y_L$. A dominant $uh$ branching ratio can also be achieved in model implementations where $U_h$ is a part of an $SU(2)$ doublet, in the limit of $y_L \gg y_R$. Conversely, the regions of parameter space where $y_R \ll y_L$ (in the case of SU(2) singlet) and $y_R \gg y_L$ (in the case of SU(2) doublet) would result in significant branching ratios to other final state such as $Zj$ and $Wj$. 

Note that most of our proposal for $U_h$ searches  (with the exception of our $b$-tagging strategy which would have to be modified) in the following sections can be applied to $Zj$ and $Wj$ final states as well, as the final state kinematics are most affected by the mass of $U_h$, and to a lesser degree by the structure of the  $U_h \rightarrow Xj$ vertex, where $X = h, W, Z$.

\section{Searching for light quark partners at the LHC Run II}\label{sec:results}
In the benchmark model we consider, the singlet partner $U_h$  decays exclusively into a Higgs and an up-type quark. The topology of  signal events  is characterized by a pair of boosted Higgs bosons (if the mass of the singlet partner is sufficiently heavy) accompanied by two light jets. We further require that the Higgs decays into $b\bar{b}$ in order to avoid a reduction of signal cross section due to small branching ratios of the Higgs to other SM final states.  Due to the boosted  Higgs topology, the final state  $b\bar{b}$ pairs are expected to be collimated into a cone of roughly $2m_h / p_T, $ where $p_T$ is the transverse momentum of the decaying Higgs. 

\begin{figure}[t]
\begin{center}
\includegraphics[scale=0.3]{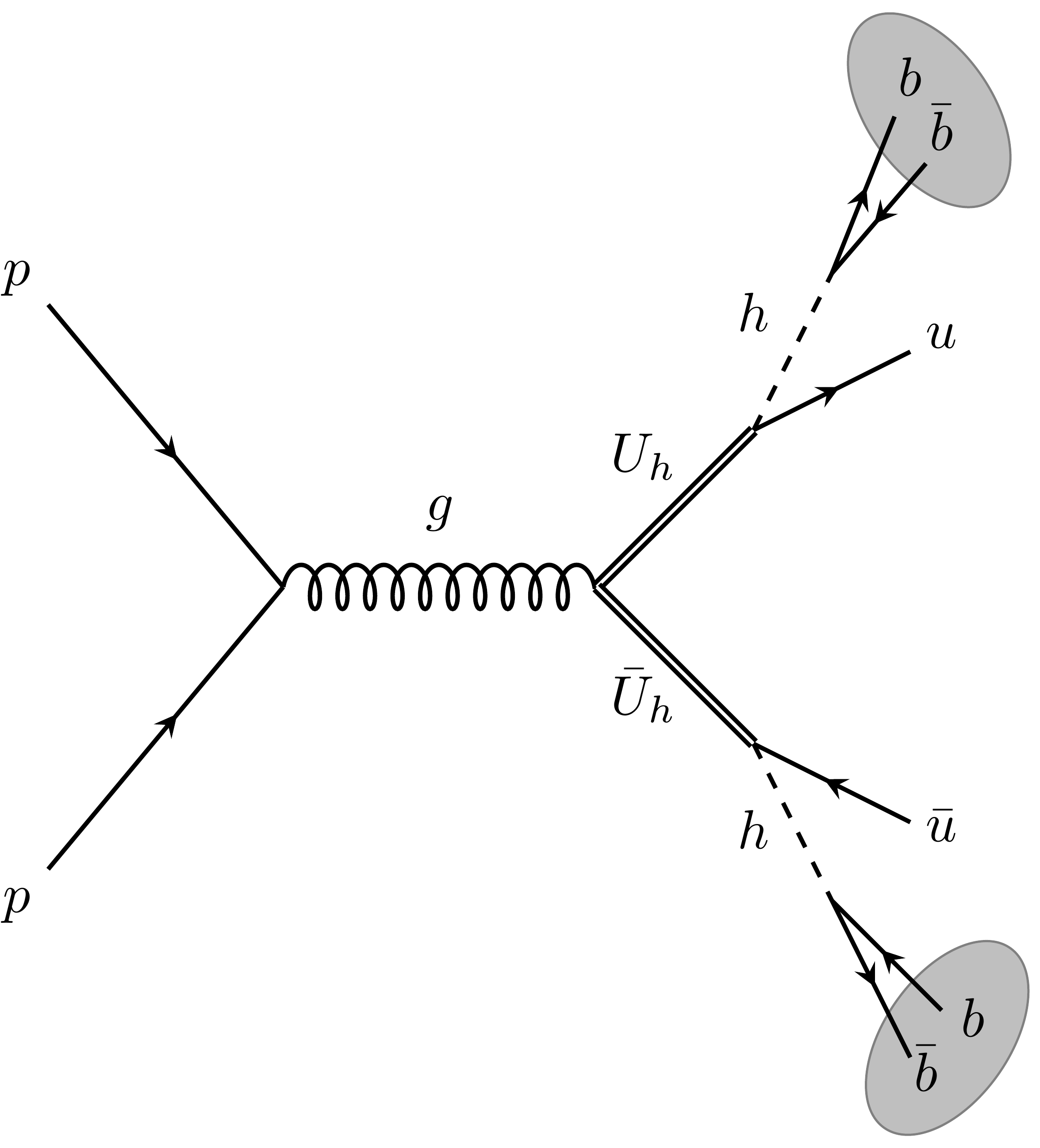}
\end{center}
\caption{The pair production channel of the $U_h$ up quark partners. Note that for $M_{U_h} \sim 1 $ TeV, the Higgs bosons are boosted, resulting in a 2 ``fat jet'' - 2 light jet event topology. 
 }
\label{fig:channels}
\end{figure}

Here we consider only pair production of $U_h$ partners at a $\sqrt{s} = 14$ TeV $pp$ collider (see Fig.\ref{fig:channels}), where the $U_h$ pairs are produced via QCD interactions. Hence, the production cross section is rather model independent, depending solely on  $M_{U_h}$. The dominant background channels to the all hadron final states in our signal events are  $t \bar{t}$ + jets, $b \bar{b}$~+~jets, and light multi-jet channels.\footnote{Another potentially interesting and very clean search channel for di-Higgs production is the di-photon $+b\bar b$ channel. However, for strongly boosted di-Higgses, the backgrounds can be efficiently removed as we will show, such that at high boost, the all hadronic channel can dominate. A qualitatively similar behavior can already been seen at both ATLAS \cite{Aad:2014xxy,ATLAS:2014xxx} and CMS \cite{CMS:2014eub,CMS:2013eua} when comparing the respective di-photon $+b\bar{b}$ and $4b$ searches at 8 TeV.}
The scope of our current effort is to study the ability of various jet observables to suppress the before-mentioned background channels and enhance the signal for $U_h$ partners of mass $~O(1 \TeV)$. To our knowledge, such searches for light quark fermionic light quark partners in the fully hadronic channels have not been studied in the past. As here we are interested in a ``proof of concept'' type of study, we will only consider signal and background events in a pileup-free environment. 

\subsection{Data Generation and Pre-Selection Cuts}
\label{sec:Basic}


We generate all events using leading order \verb|MadGraph 5| \cite{Maltoni:2002qb} at a $\sqrt{s} = 14$ TeV $pp$ collider, assuming a  CTEQ6L \cite{Nadolsky:2008zw} set of parton distribution functions. At the hard process level, we require that all final state partons pass cuts of $p_T > 15$~GeV,~$| \eta |< 5$. Next, we shower the events with \verb|PYTHIA 6| \cite{Sjostrand:2006za} using the MLM-matching scheme \cite{Artoisenet:2010cn} with \verb|xqcut|~$> 20$~GeV and \verb|qcut|~$> 30$~GeV. We match the multi-jet events up to four jets, while the $t\bar{t}$ and $b\bar{b}$ samples are matched up to two extra jets. We cluster all showered events with the \verb|fastjet| \cite{Cacciari:2011ma} implementation of the anti-$k_T$ algorithm \cite{Cacciari:2008gp}.


In order to perform the analysis with a manageable number of events in the background channels ($i.e. \, \sim 10^6 $), we impose a generator level cut on $H_T$, a scalar sum of all final state parton transverse momenta. The motivation for the generator level $H_T$ cut comes from the fact that pair produced light quark partner events contain two objects of mass $\sim 1 $ TeV, implying that the signal will be characterized by $H_T$ of roughly 2 TeV. In order to avoid possible biases on the background data by increasing the $H_T$ cut too much, we hence require $ H_T > 1.6$ TeV on all generated backgrounds. 

\begin{figure}[t]
\includegraphics[scale=0.5]{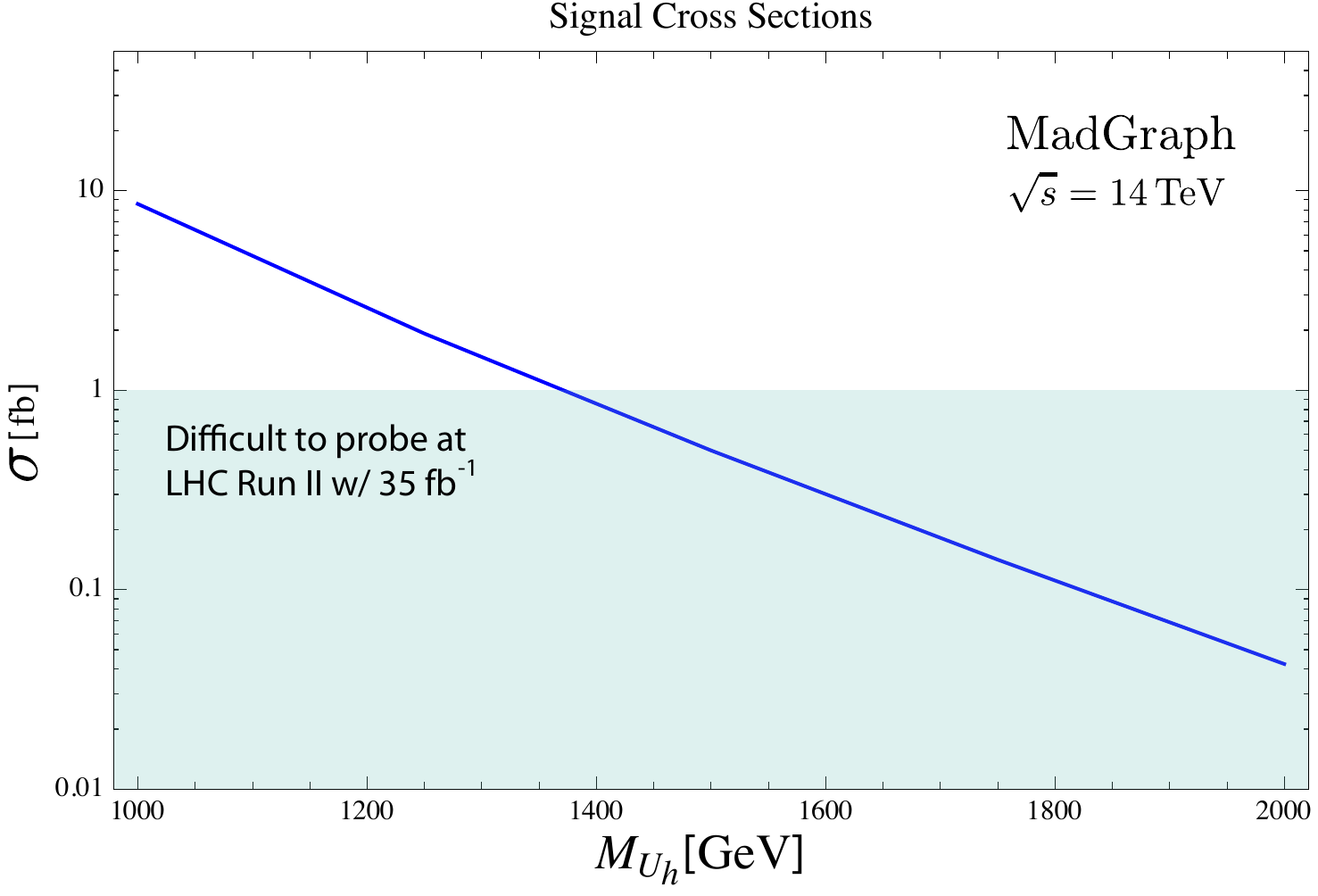}
\includegraphics[scale=0.5]{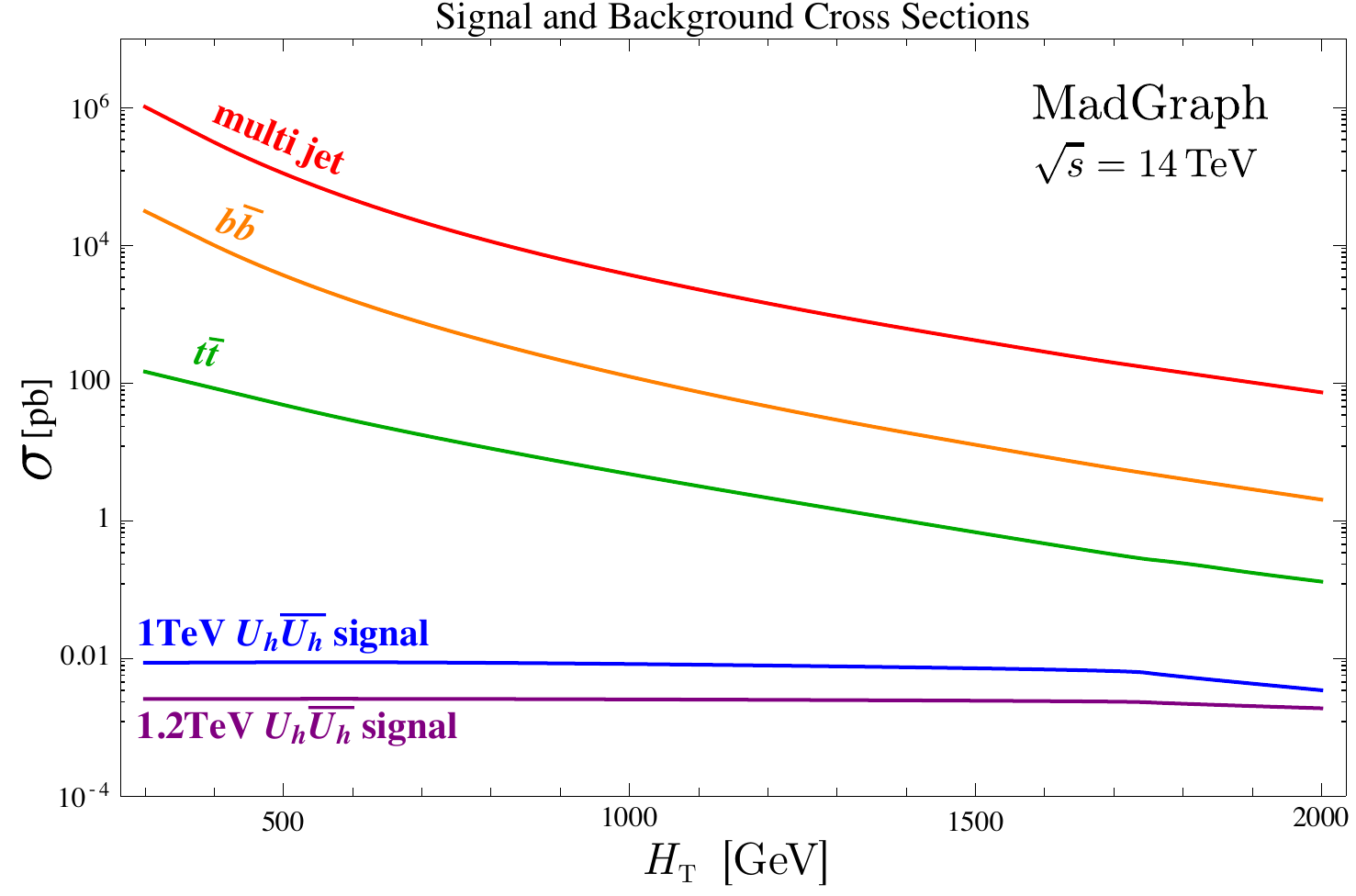}
\caption{Left: The cross section of the signal events as a function of $M_{U_h}$ with the basic pre-selection cuts, $p_T > 15$~GeV, $|\eta| < 5$ and  $H_T > 0$ GeV. \newline Right: Signal and background cross sections as a function of $H_T$ cut. The plot is normalized to NLO as in Table \ref{tab:HT1600}.}
\label{fig:Partonic2}
\end{figure}

We summarize the cross sections for the signal parameter point of $M_{Uh} = 1 \TeV$ and the most dominant backgrounds in Table~\ref{tab:HT1600}. For completeness, we show the $U_h$ pair production cross section as function of $M_{Uh}$ in Fig.\ref{fig:Partonic2}, where we assume ${\rm Br}(U_h \rightarrow hu) = 1$ and the branching ratio of Higgs to a pair of $b$ quarks is included. Notice that  the total production cross section for partner masses above 1.3 TeV goes into  the sub-femtobarn region which will be challenging to probe at the Run II of the LHC with $35 \fb^{-1}$ of integrated luminosity.
A closer look at the numerical values of the signal and background cross sections suggests that a total improvement in $S/B$ of  $\mathcal{O} (10^5)$ is desired to reach $S/B \sim 1$. For that purpose, we will introduce a  new cut scheme in Section~\ref{sec:cutflow}, which exploits the characteristic topology and kinematic features of the signal events.


\vspace{1cm}
\begin{table}[h]
\begin{center}
\begin{tabular}{|c|c|c|c|}
\hline
	$\sigma^{\rm LO}_{s}$ [pb]   &  $\sigma^{\rm NLO}_{t\bar{t}}$ [pb]	&   $\sigma^{\rm NLO}_{b\bar{b}}$ [pb]  & $\sigma^{\rm NLO}_{\mathrm{multi-jet}}$ [pb]	\\ \hline
     $6.8 \times 10^{-3}$		& 	$4.6 \times 10^{-1}$		&  	8.4				&  	282.2			 \\  \hline		
\end{tabular}\par
\end{center}
\caption{Cross sections for the $U_h \bar{U}_h$ pair production (assuming $M_{U_h} = 1 \TeV$) and backgrounds (assuming $H_T > 1600$ GeV), at 14 TeV LHC. We normalize the `` $t \bar{t}$ +0,1,2 jets " to the NNLO + NNLL result of Ref.  \cite{Czakon:2013goa}, while for the rest of the backgrounds we use a conservative estimate for the NLO K-factor of 2.0.}
\label{tab:HT1600}  
\end{table}

\subsection{Tagging of Boosted Higgs Jets}
\label{sec:TOM}
The decay products of a boosted Higgs are collimated into a cone of  $R \sim 2m_h / p_T$, where $p_T$ is the transverse momentum of the Higgs boson. Since we consider light quark partners of mass $\sim 1 \TeV$, the resulting Higgs bosons will have $p_T \sim 500$ GeV, and hence will decay into a cone of roughly $R \sim 0.5$. Clustering the decay products of a boosted Higgs into a large cone ($e.g.  \, R=0.7$), will typically result in a single ``fat jet'' of mass $\sim m_h$. However, traditional jet observables such as jet  $p_T$ and $m$ are inadequate to efficiently  distinguish between Higgs, top and QCD ``fat jets'', and a further consideration of Higgs ``jet substructure,'' is needed to reduce the enormous QCD backgrounds. 
Many methods designed to tag the characteristic ``two prong'' substructure of the hadronically decaying Higgs exist in the literature \cite{Butterworth:2008iy, Almeida:2011aa, Backovic:2012jj, Schlaffer:2014osa, Soper:2011cr}. Here we will use the \verb| TemplateTagger v.1.0 |\cite{Backovic:2012jk} implementation of the Template Overlap Method \cite{Almeida:2010pa, Almeida:2011aa,  Backovic:2012jj, Backovic:2013bga}. 

The Template Overlap algorithm for boosted jet tagging attempts to match a parton level model (template) for a boosted jet decay  ($i.e.$ the $b \bar{b}$ system with the constraint of 
$( p_1 + p_2 )^2= m_h^2  $ ) to the energy distribution of a boosted jet. The procedure is performed by minimizing the difference between the calorimeter energy depositions within small angular regions around the template patrons and the parton energies, over the allowed phase space of the template  four-momenta. Refs. \cite{Almeida:2010pa, Almeida:2011aa,  Backovic:2012jj} studied the use of TOM to tag boosted Higgs decays in the context of the Standard Model. To our knowledge, our current effort is the first attempt to utilize TOM for boosted Higgs studies in a BSM scenario. 

An attractive feature of TOM is a relatively weak susceptibility to pileup contamination \cite{Backovic:2013bga}. The overlap analysis is affected only by the calorimeter depositions which land in angular regions of typically $ r \sim 0.1$ from the template patrons. The rest of the jet energy distribution does not contribute to the estimates of the likelihood that a particular template matches the jet energy distribution. As pileup contamination scales as $R^2$, where $R$ is the jet cone, the effects of pileup on the TOM analysis will be of order few percent, compared to (say) the $p_T$ of a typical fat jet of $R \sim 1.0.$  

Ideally, in order to maximize the information extracted from jet substructure, one would perform TOM analysis for all heavy standard model decays on each candidate fat jet. Such analysis would result in a vector of overlap scores 
\begin{equation}
\overrightarrow{Ov} = (Ov_2^i; Ov_3^i)\,,
\end{equation}

where $i = W, Z, h, t$. Various correlations within the multi-dimensional overlap space could then be exploited to fully maximize the ability of TOM to tag the desired heavy particles. 
The full multi-dimensional TOM analysis is beyond the scope of our current effort and we find it sufficient to use only  a combination of two body Higgs as well as three body top template analysis (in order to further suppress the large $t\bar{t}$ background) \footnote{Note that the addition of a three body (NLO) Higgs template analysis could further suppress the multi-jet and $b\bar{b}$ backgrounds, but would not significantly help in suppression of the $t\bar{t}$ background \cite{Backovic:2012jj}.}. As the three prong decay of a boosted top is more complex of an object than the typical two prong decay of a boosted Higgs, it is possible for a top fat jet to pass the two-body Higgs template tagging procedure. On the other hand, it is difficult for a Higgs to appear as a fake top \cite{Backovic:2012jj}. We hence require all Higgs candidate jets to pass the requirement 
\begin{equation}
	Ov_2^h > 0.4, \,\,\,\,\, Ov_3^t <0.4 \,.
\end{equation}
As we will show in the following sections, the combined requirement on $Ov_2^h$ and $Ov_3^t$ is very efficient at removing the $t\bar{t}$ fake rate.

For the purpose of this analysis, we generate 17 sets of both two body Higgs and three body top templates at fixed $p_T$, starting from $p_T = 425 \GeV$ in steps of $50 \GeV$, while we use a template resolution parameter $\sigma = p_T /3$ and scale the template subcones according to the rule of Ref. \cite{Backovic:2012jj}.

\subsection{$b$-tagging}
\label{sec:b-tagging}

The signal final states we consider contain four $b$-jets from two Higgses, which can be extremely useful in disentangling the signal events from the background channels. However, requiring four $b$-tags in a boosted configuration comes at a severe cost of the signal efficiency as even in the optimistic scenario of a single $b$-tag efficiency of 75\%, $b$-tagging four jets alone would cut out about 70\% of the signal events. Instead, here we will consider two $b$-tags, and require that they are contained within the two Higgs candidate jets.

A full analysis of $b$-tagging requires a detailed detector study which is beyond the scope of our work. Here we adopt a simplified, semi-realistic $b$-tagging procedure, whereby we assign to each $r=0.4$ jet a $b$-tag if there is a parton level $b$ or $c$ quark within $\Delta r=0.4$ from the jet axis. We then weight each event by the benchmark $b$-tagging efficiencies:
\begin{equation}
	\epsilon_b = 0.75, \,\,\,\, \epsilon_c = 0.18, \,\,\,\,\, \epsilon_j = 0.01\,, 
\end{equation}
where $\epsilon_{b, c, j}$ are the efficiencies that a $b$, $c$ or a light jet will be tagged as a $b$-jet. 
For a Higgs fat jet to be $b$-tagged, we then require that a $b$-tagged $r=0.4$ jet lands within $\Delta R=0.7$ from the fat jet axis. Furthermore, we take special care of the fact that more than one $b$-jet might land inside the fat jet and reweigh the $b$-tagging efficiencies according to the rule of Table \ref{tab:btag}.

\vspace{1cm}
\begin{table}[h]
\begin{center}
\begin{tabular}{|c|c|c|}
\hline
	$b$-tag scores of a fat jet	 &  Efficiency & values \\ \hline
 	 0 (jet: u,d,s,g)				& 	 	$\epsilon_j$				&  0.01	\\  \hline
	1 (1c)					& 	 	$\epsilon_c$					&  0.18	\\  \hline
	2 (2c)					& 	 	$2\, \epsilon_c (1-\epsilon_c)+ {\epsilon_c}^2$		&  0.33			\\  \hline
	3 (1b)					& 	 	$\epsilon_b$					&  0.75		\\  \hline
	4 (1b+1c)					& 	 	$\epsilon_b (1-\epsilon_c) + \epsilon_c (1-\epsilon_b)+\epsilon_b \epsilon_c$		& 	0.80			\\  \hline
	5 (1b+2c)					& 	 	$\epsilon_b (1-\epsilon_c)^2 + 2 (1-\epsilon_b) (1-\epsilon_c) \epsilon_c + \epsilon_b {\epsilon_c}^2$	& 0.60	\\  \hline
	6 (2b)					& 	 	$2 \epsilon_b (1-\epsilon_b)+ {\epsilon_b}^2$				&   0.94		\\  \hline
	7 (2b+1c)					& 	 	$1 - (1-\epsilon_c) (1-\epsilon_b)^2$				&   0.95		\\  \hline
	8  (2b+2c)					& 	 	$1 - (1-\epsilon_c)^2 (1-\epsilon_b)^2$				& 	0.96	\\  \hline
	9  (3b)					& 	 	$1 - (1-\epsilon_b)^3$						&  0.98	\\  \hline					 
\end{tabular}\par
\end{center}
\caption{Efficiency that a Higgs fat jet will be $b$-tagged assuming that it contains a specific number of light, $c$ or $b$ jets within $\Delta R = 0.7$ from the jet axis. $\epsilon_j$, $\epsilon_c$ and $\epsilon_b$ are $b$-tagging efficiencies for light, $c$ and $b$ jets respectively. We neglect the possibilities beyond three proper $b$-tagged jets within a fat jet as they occur at  too low of a rate to be significant. } \label{tab:btag} 
\end{table}

\subsection{Event Selection and Reconstruction of the $U_h$ Pair}\label{sec:cutflow}

We proceed to discuss in detail the cut scheme we propose for the all-hadronic searches for pair produced $U_h$ partners. For the convenience of the reader, we outline the event selection in Table \ref{tab:Cuts1},  while a detailed  description  and definition of the observables  can be found in the following text.

\begin{table}[h]
\begin{center}
\begin{tabular}{|c|c|c|}
\hline
\multirow{9}{*}{Cut Scheme}	 &   \multirow{7}{*}{Basic Cuts}   	& Demand at least four fat jets ($R = 0.7$)  with  		 \\   
						&							& $p_T > 300$ GeV, $| \eta |< 2.5$ 		\\ \cline{3-3}
			           &      	& Declare the two highest $p_T$ fat jets   \\  
			           &		& satisfying $Ov_{2}^h > 0.4$ and $Ov_{3}^t < 0.4$\\ & &  to be Higgs candidate jets.		\\ 
			            &        &   At least 1$b$-tag on both Higgs candidate jets. \\ \cline{3-3}
			           &		& Select the two highest $p_T$ light jets ($r=0.4$) \\
			           &		&  with $p_T > 25 \GeV$ to be the $u$ quark candidates.    \\  \cline{2-3}
 				   &  \multirow{4}{*}{Complex Cuts}       &    $|\Delta_{h}  | < 0.1$ 		 \\  
	 		 	 &         &  $| \Delta_{U_h} | < 0.1$		 \\  
				   &        &    $m_{U_{h1,2}} > 800$ GeV			 \\  
				   	 \hline
		 				
 	\end{tabular}\par
	\end{center}	
\caption{Summary of the Event Selection Cut Scheme. } \label{tab:Cuts1} 
\end{table}

\begin{figure}[h]
\includegraphics[scale=0.35]{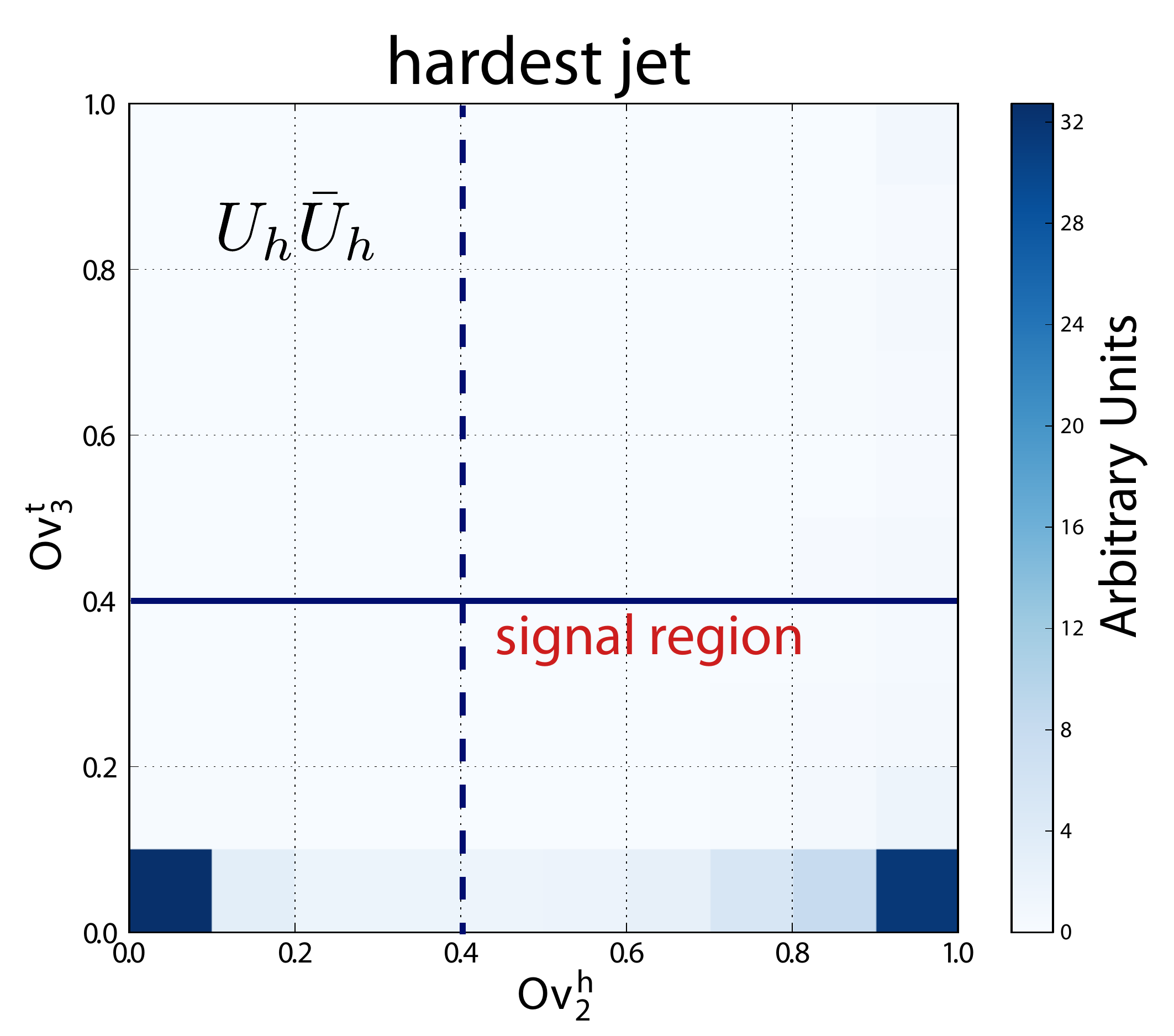}
\includegraphics[scale=0.35]{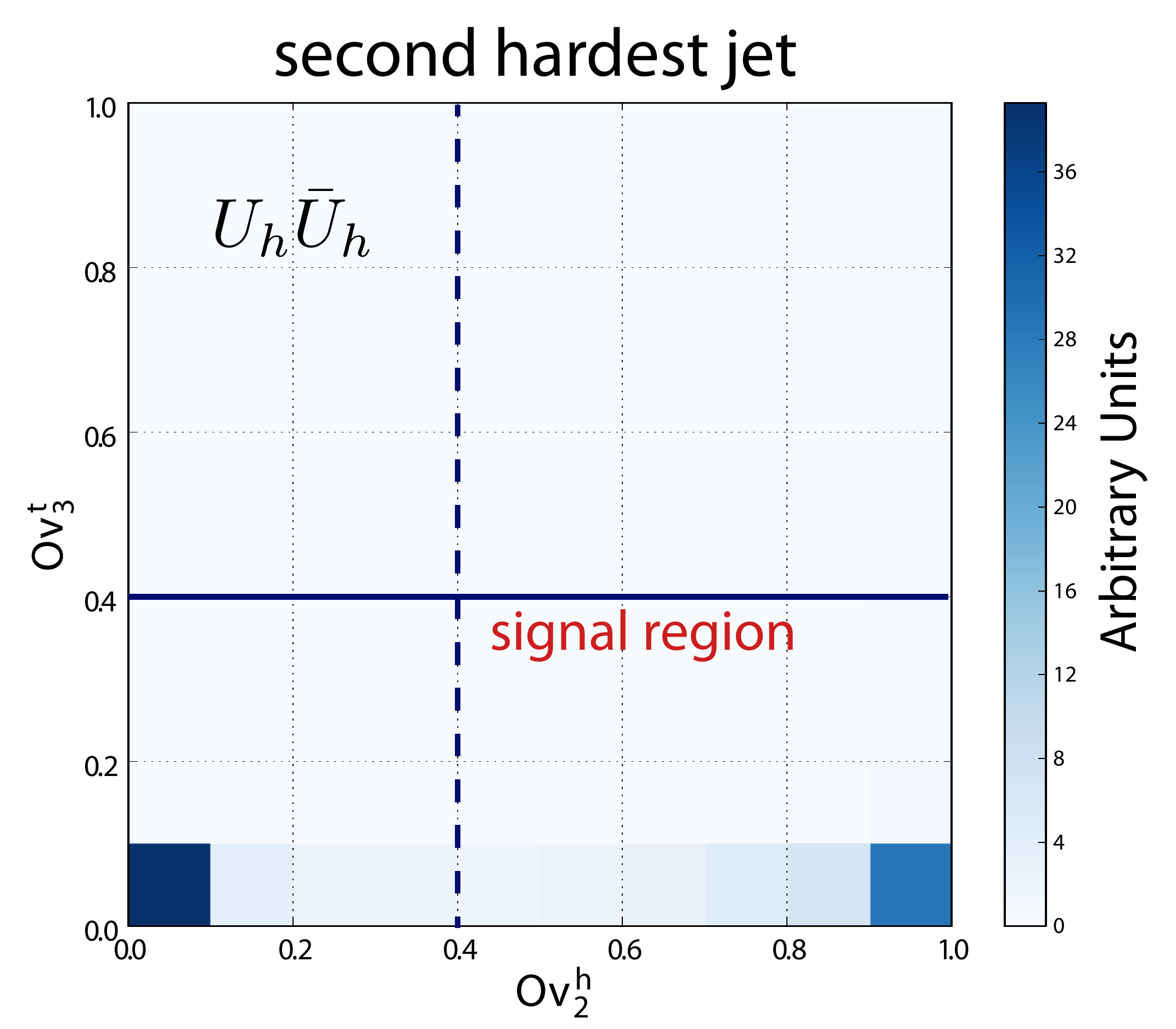}
\includegraphics[scale=0.35]{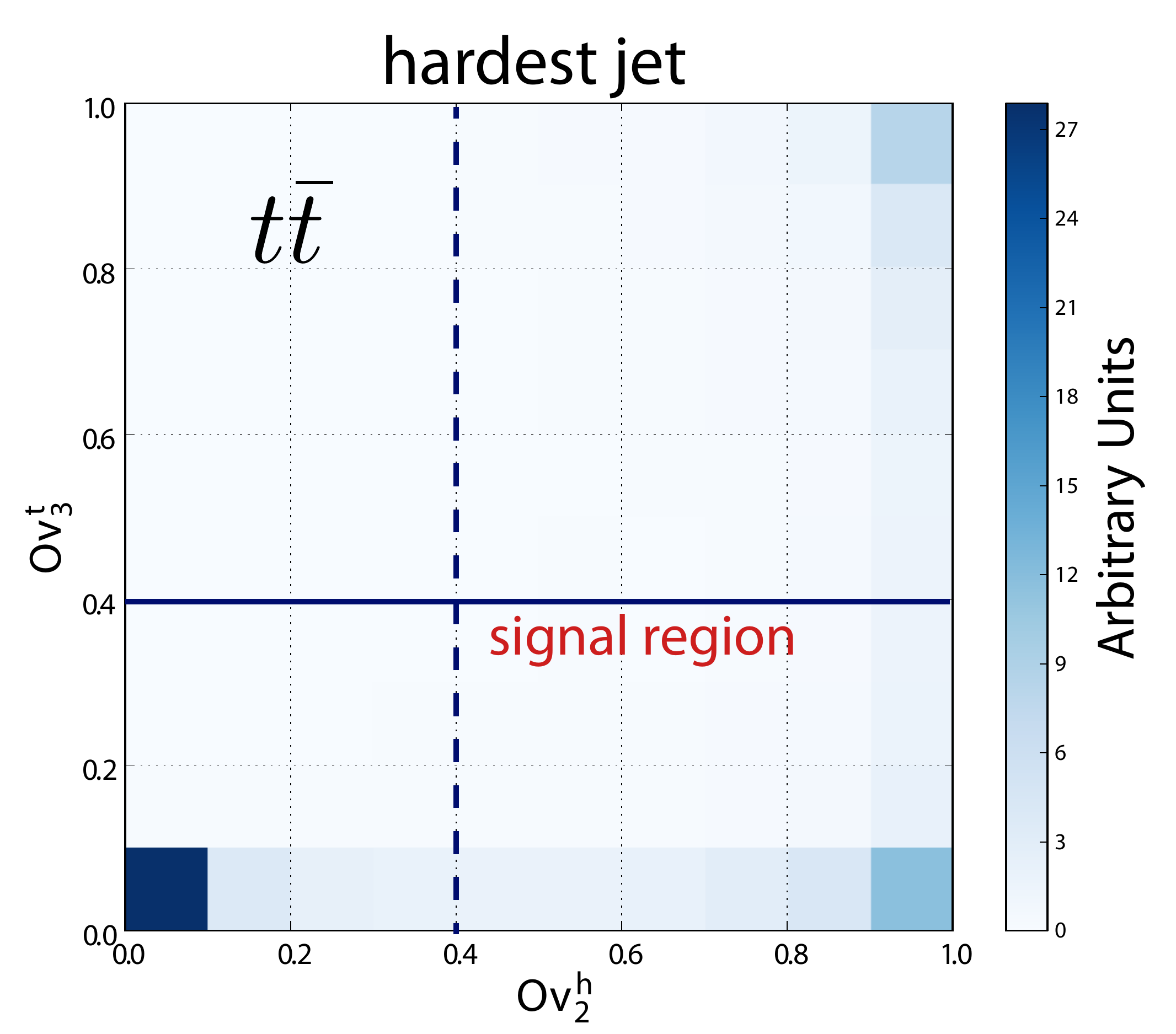}
\includegraphics[scale=0.35]{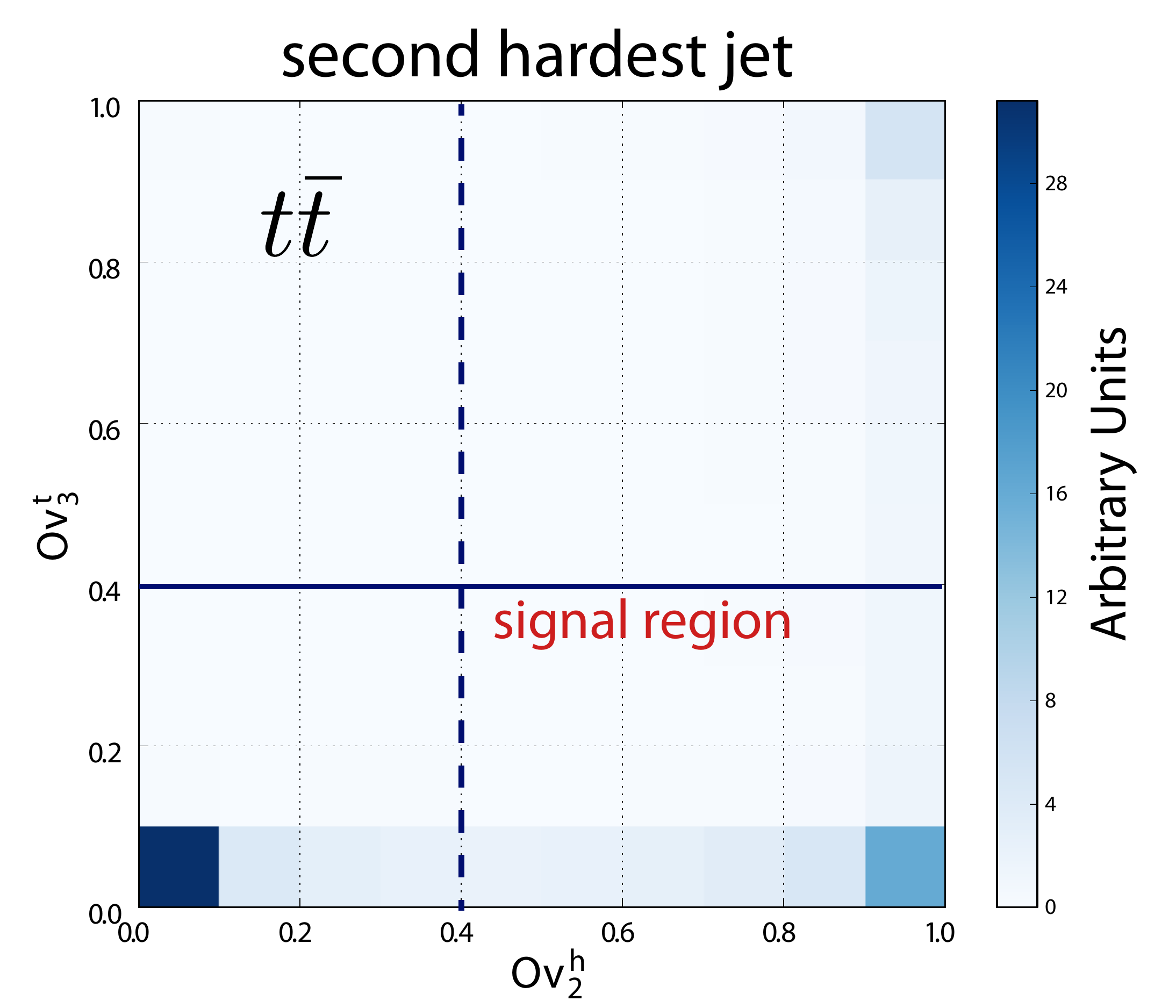}
\caption{Two dimensional distributions of peak template overlap scores for the two highest $p_T$ fat jets in the event. The top panels show the events from pair produced $U_h$, while the bottom shows the $t\bar{t}$ background. Here we omit showing $b\bar{b}$ and multi-jet backgrounds, as their overlap distributions are trivial and simply peak at $Ov_2^h \approx Ov_3^t\approx 0$.}
\label{fig:Overlaps2d}
\end{figure}

We begin by requiring at least four anti-$k_T$, $R=0.7$ jets with 
\begin{equation}
	p_T^{R=0.7} > 300 \GeV, \,\,\,\,\,\,\, |y^{R=0.7}| < 2.5\,.
\end{equation}
The requirement on the presence of four fat jets pre-selects signal event candidates, as we expect two pairs of boosted Higgs-light jets to appear in the final state \footnote{Selecting 4 $R=0.7$ fat jets also simplifies the TOM jet substructure analysis. }. In order to determine which of the four jets are the Higgs candidates, we select the two highest $p_T$ fat jets which satisfy the TOM requirement of 
\begin{equation}
	Ov_2^h > 0.4, \,\,\,\,\,\,\,\, Ov_3^t < 0.4 \,, \label{eq:tom_reqs}
\end{equation}
of Section \ref{sec:TOM}. The requirement on peak template overlap is designed to select the two Higgs candidate jets in the event, while ensuring that the jets are not fake tops. If less than two fat jets pass the overlap requirement, the event is rejected.

The overlap selections in Eq.~\eqref{eq:tom_reqs} deserve more attention. Figure \ref{fig:Overlaps2d} illustrates how utilizing multi-dimensional TOM analysis ($i.e. \, Ov_2^h$ and $Ov_3^t$) can help in reducing the background contamination of signal events. If we consider only $Ov_2^h$ (dashed line), a significant fraction of $t\bar{t}$ would pass any reasonable overlap cut. However, in a two dimensional distribution, it is clear that many of the $t\bar{t}$ events which obtain a high $Ov_2^h$ also obtain a high $Ov_3^t$ score. Contrary to $t\bar{t}$ events, the signal events almost never get tagged with a high $Ov_3^t$ score, as it is difficult for a proper Higgs fat jet to fake a top. Hence, an upper cut on $Ov_3^t$ (solid line) efficiently eliminates a significant fraction of $t\bar{t}$ events, at a minor cost of signal efficiency. Note that the peak at $Ov_2^h \approx Ov_3^h \approx 0$ in the signal distributions corresponds to events where the hardest/second hardest fat jet is likely a light jet.

\begin{figure}[b]
\begin{tabular}{cc}
\includegraphics[scale=0.37]{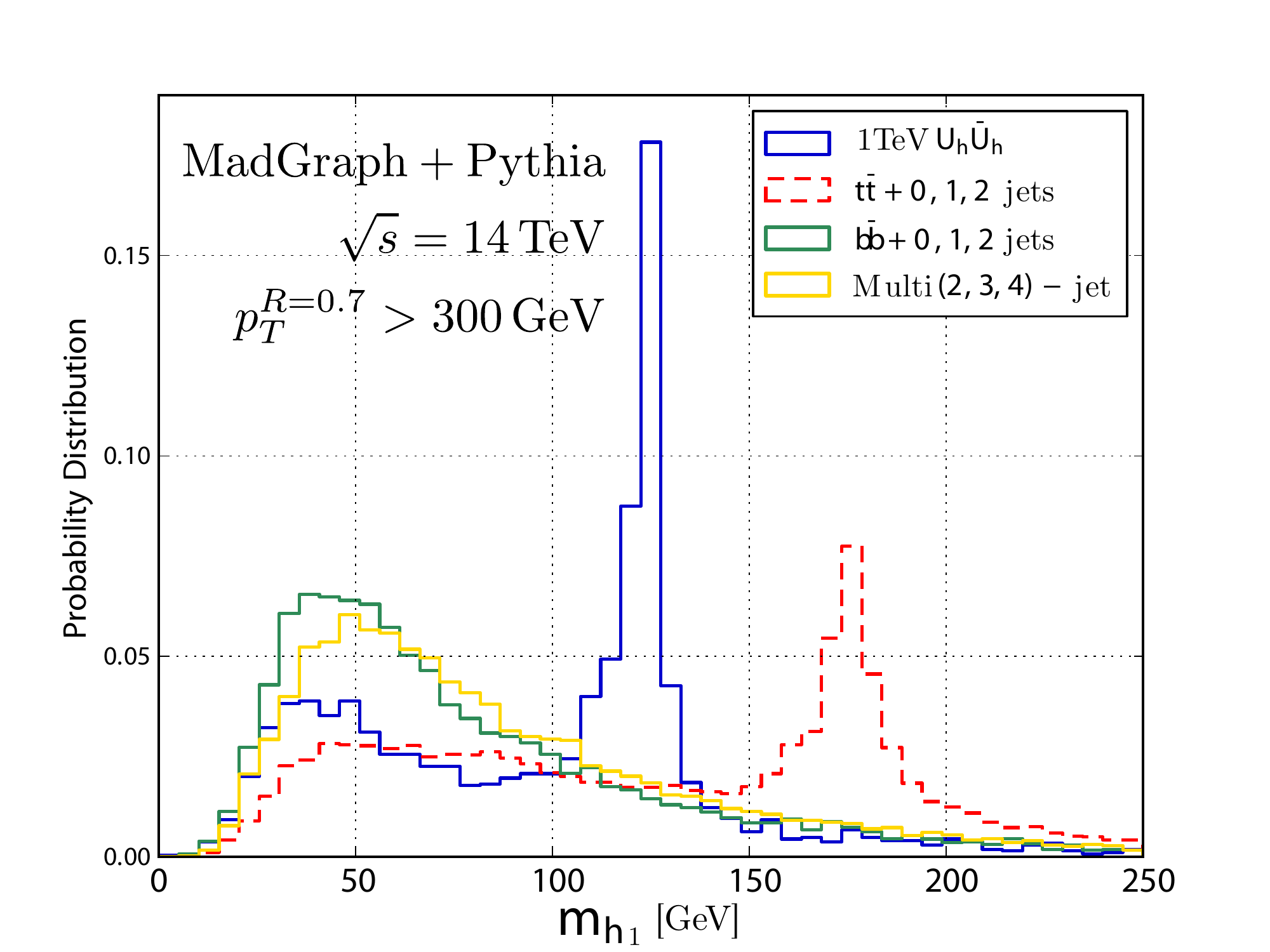}&
\includegraphics[scale=0.37]{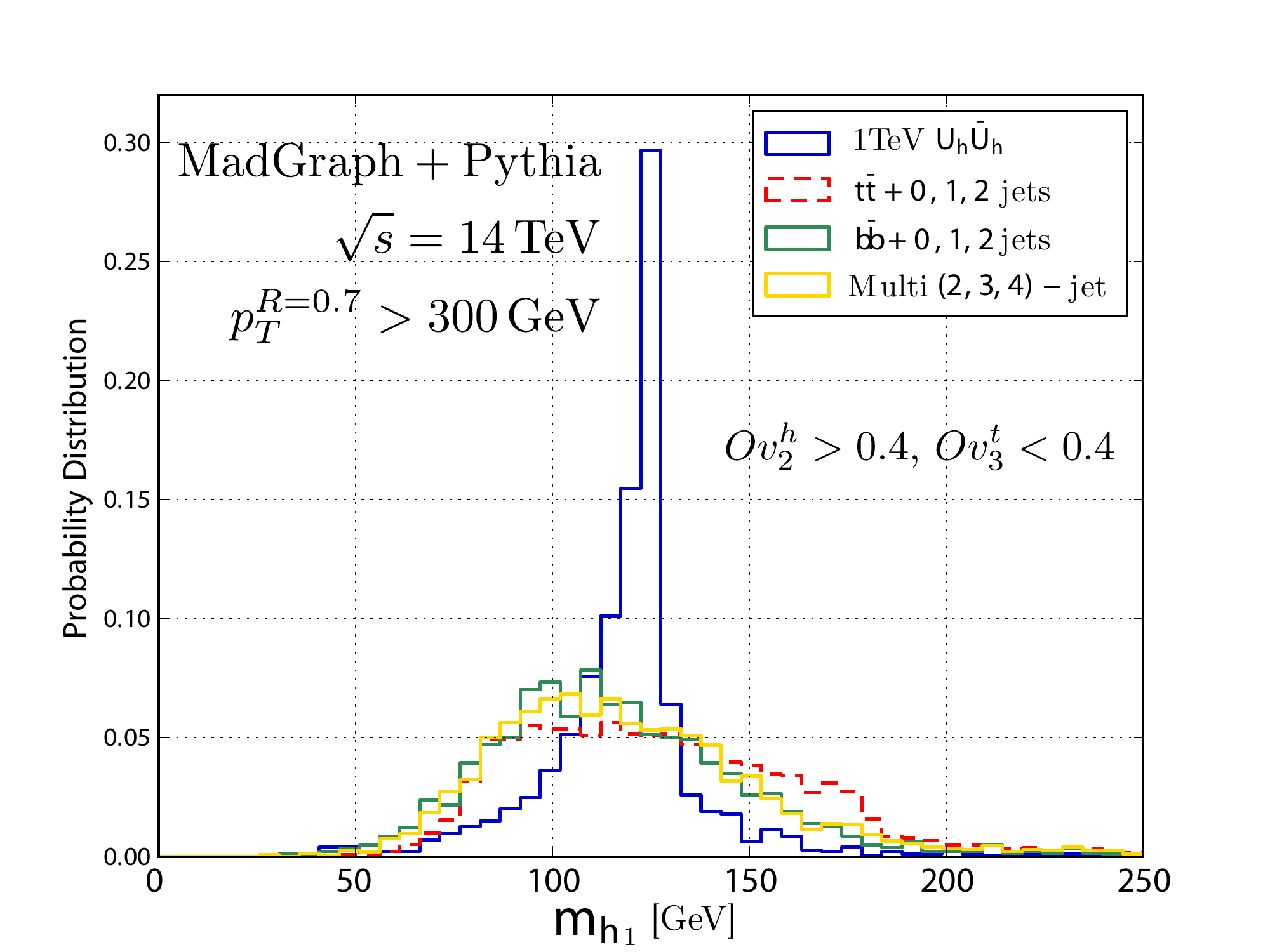}\\
\includegraphics[scale=0.37]{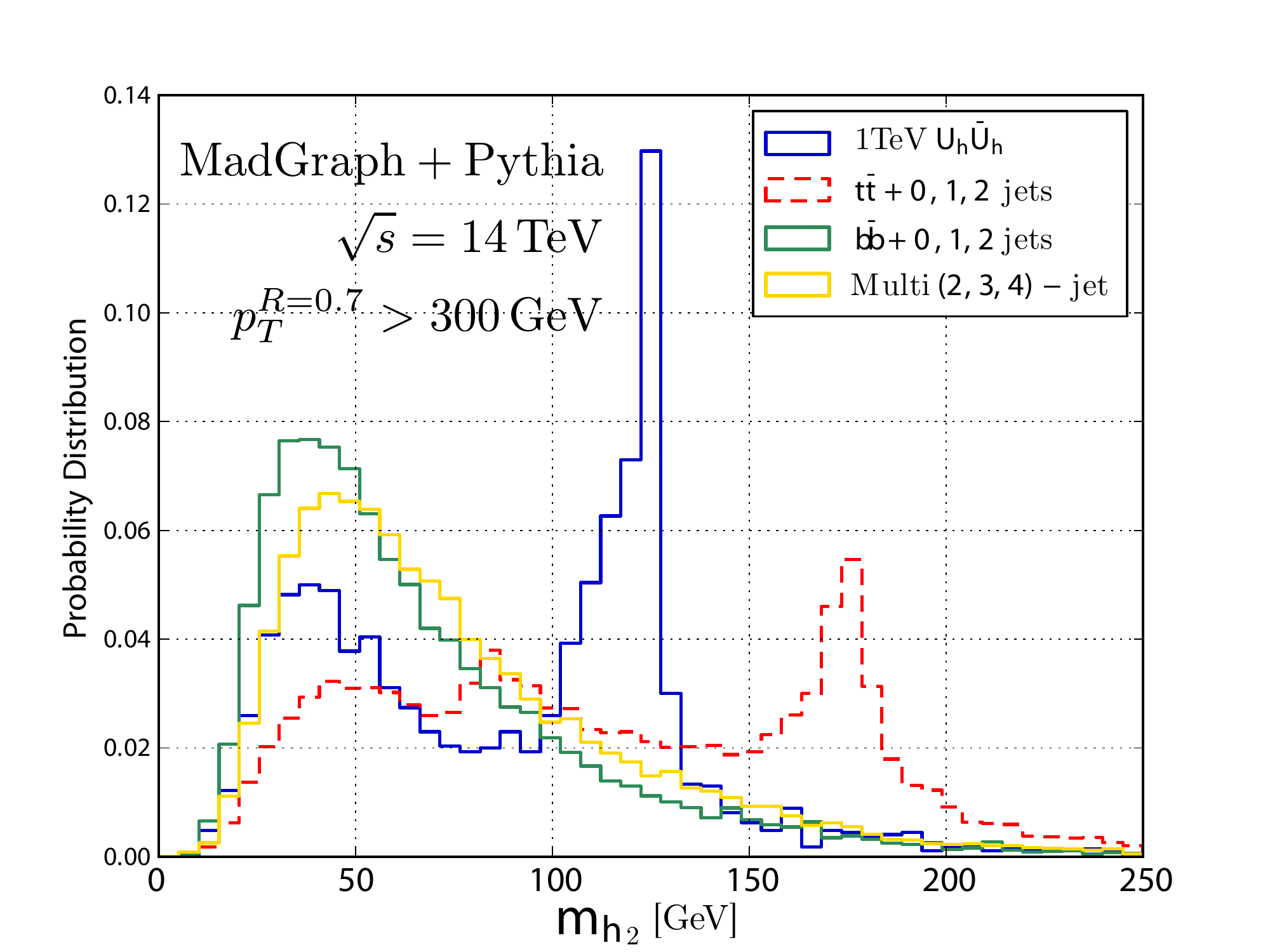}&
\includegraphics[scale=0.37]{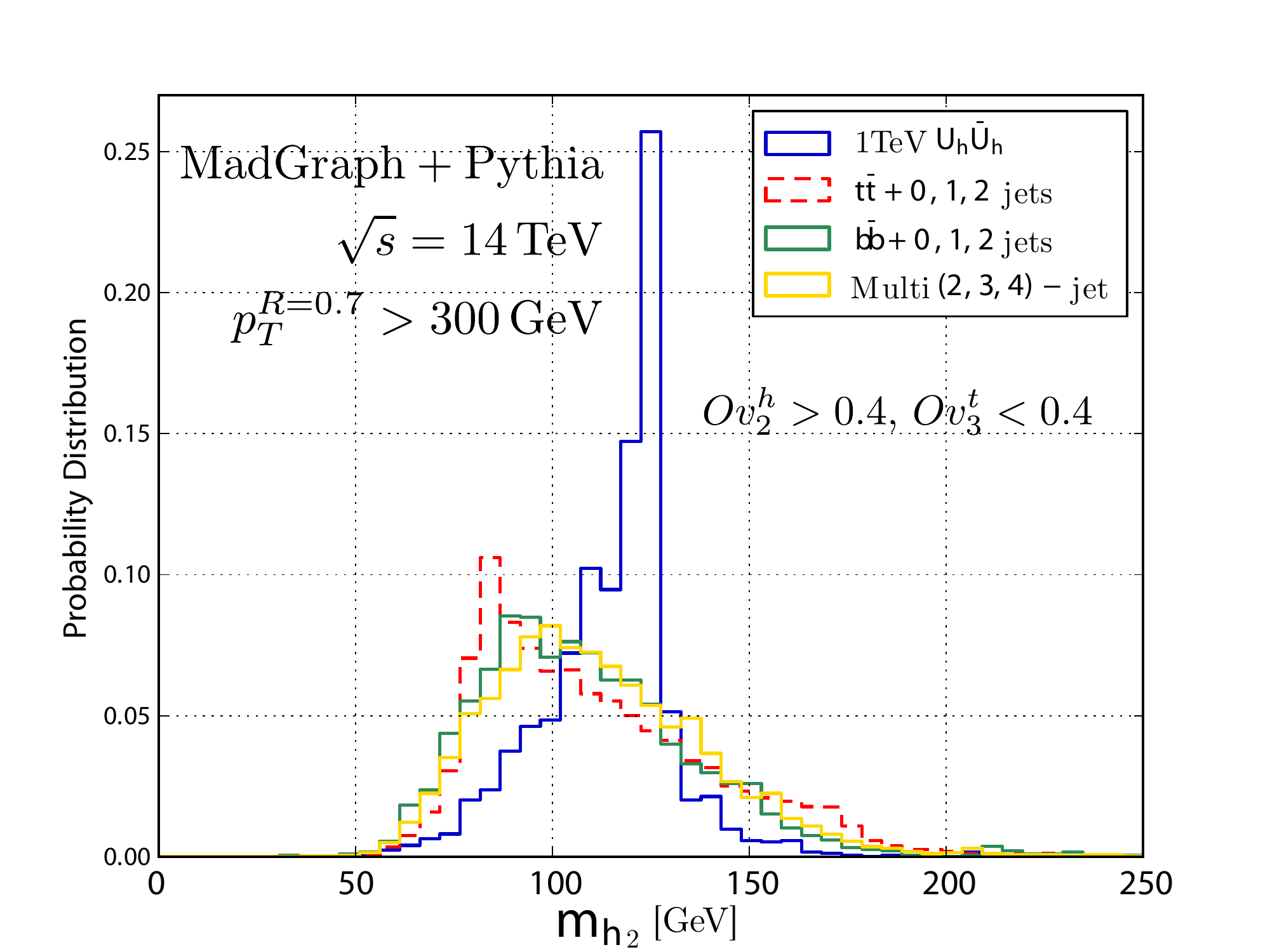}
\end{tabular}
\caption{The invariant mass of the two highest $p_T$ fat jets ($R=0.7$) (labeled $h_{1, 2}$) before (left panels) and after (right panels) the boosted Higgs selection criteria. Notice that TOM selection filters out well both the high mass and low mass background events.}
\label{fig:MassPlotsh}
\end{figure}

Figure \ref{fig:MassPlotsh} illustrates the effects of $Ov$ cuts on the mass distribution of the two highest $p_T$ jets. Note that the intrinsic mass filtering property of TOM can be clearly seen in the results. The mass resolution of the Higgs fat jets improves upon the cut on the overlap, while the contributions from both high mass and low mass background regions are significantly diminished. 

In addition to jet substructure requirements for Higgs tagging, we require both Higgs candidate jets to contain at least one $b$-tagged $r=0.4$ jet within the fat jet, as prescribed in Section \ref{sec:b-tagging}.

In order to pick out the light jets, we re-cluster each event with $r=0.4$ (also necessary for $b$-tagging) and select the two highest $p_T$ jets which pass the requirement of
\begin{equation}
 	p^{r=0.4}_T > 25 \GeV, \,\,\,\,\, |y^{r=0.4}| < 2.5, \,\,\,\,\,\, \Delta R_{uh} > 1.1\,,
\end{equation}

where $\Delta R_{uh}$ stands for the plain distance in $\eta, \phi$ between the $r=0.4$ jet ($i.e.$ the up type quark) and each of the Higgs candidate fat jets. We declare these jets to be the $u$ quark candidates.

\begin{figure}[b]
\includegraphics[scale=0.37]{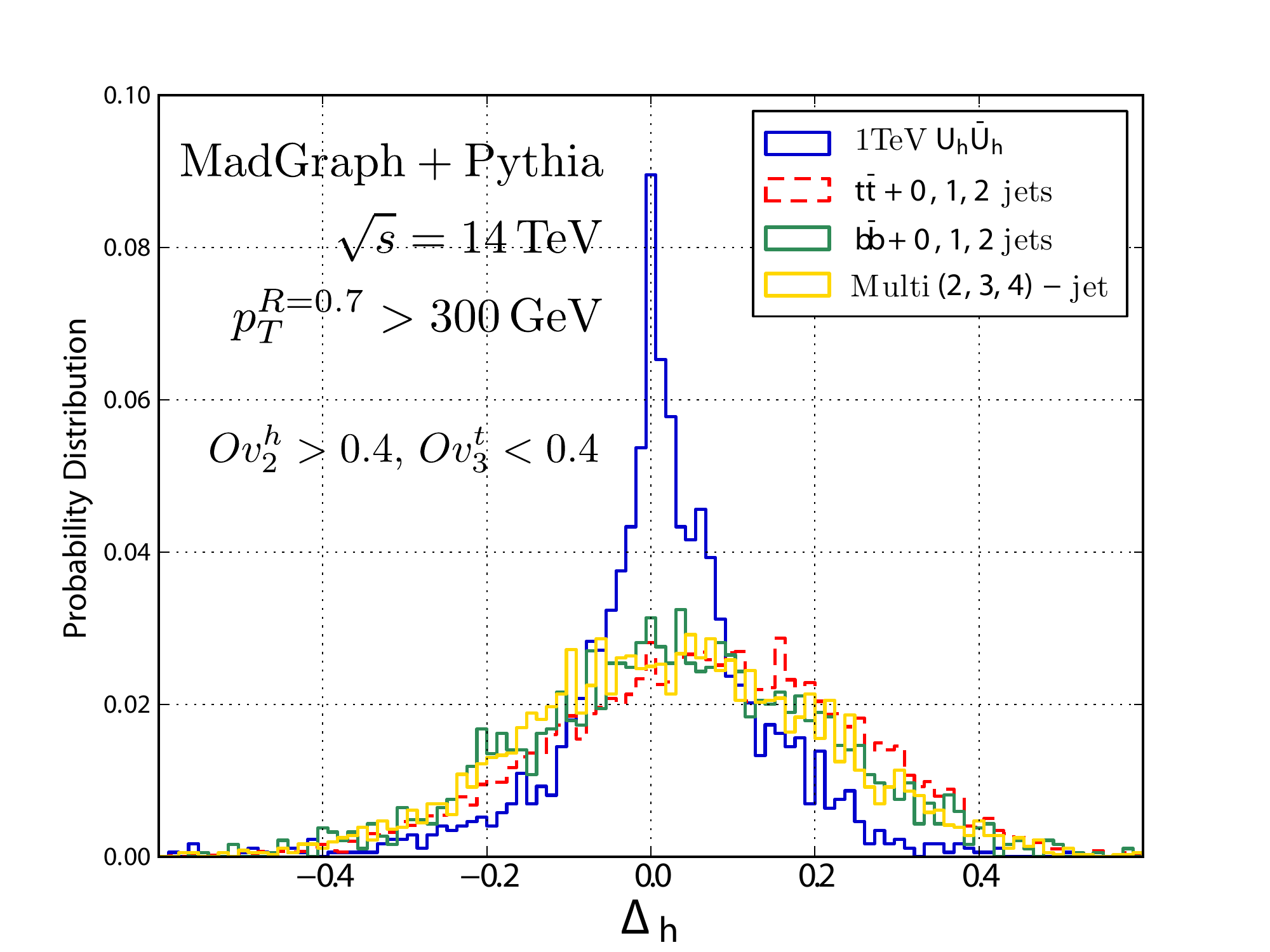}
\includegraphics[scale=0.37]{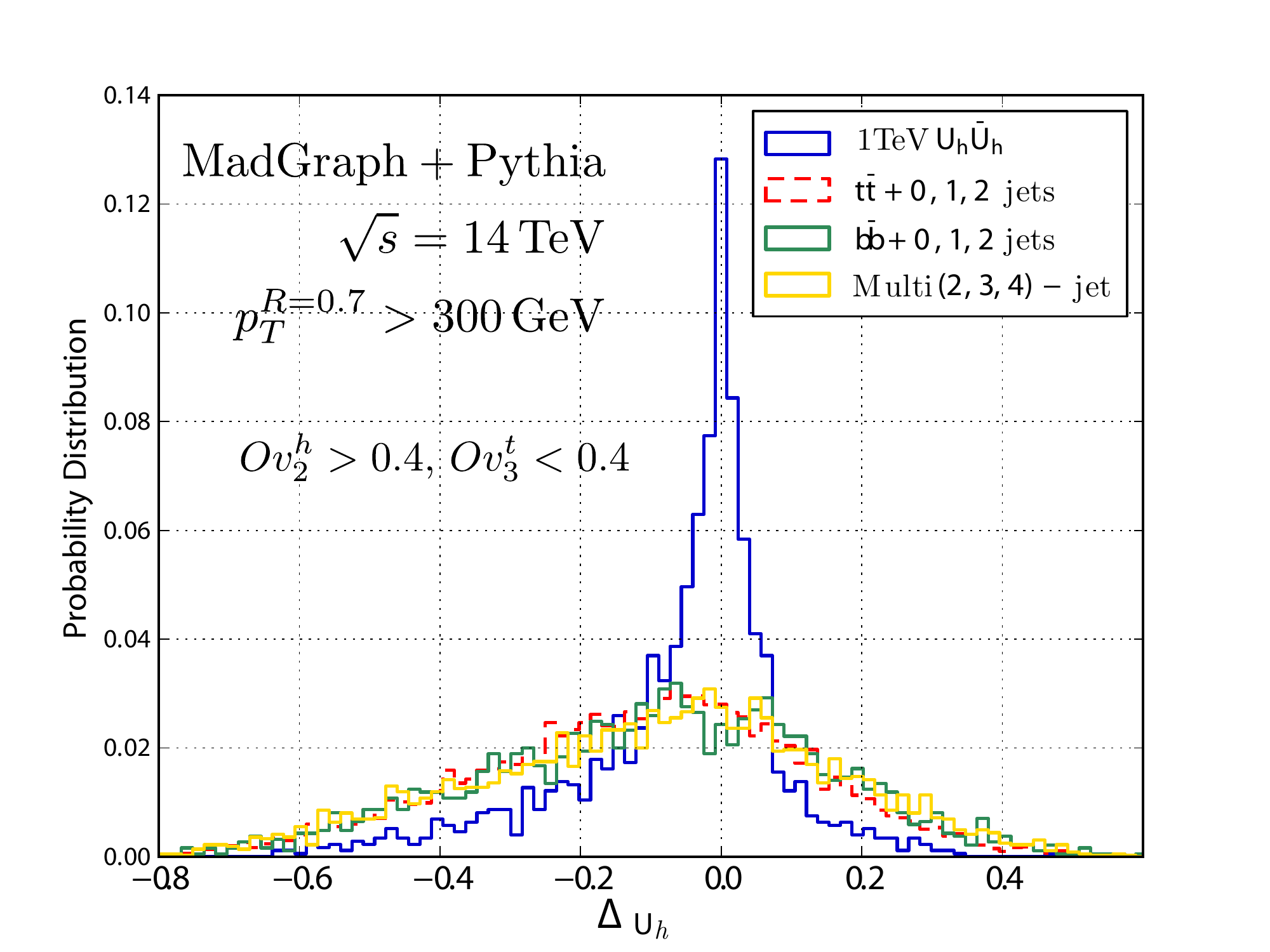}
\caption{Left panel: mass asymmetry $\Delta_h$ between the two highest $p_T$ fat jets which pass the Higgs tagging requirement.  Right panel: minimized mass asymmetry $\Delta_{U_h}$ of the reconstructed $U_h$ pair. }
\label{fig:MassPlotsDeltas}
\end{figure}

Since we expect two Higgs fat jets in the final state, a comparison between the masses of the two hardest fat jets which pass the overlap criteria provides a useful handle on the background channels. In order to exploit this feature, we construct a mass asymmetry
\begin{equation}
\Delta_{h} \equiv \frac{m_{h1} - m_{h2}}{m_{h1} + m_{h2}}\,,
\end{equation}
where $m_{h_{1,2}}$ are the masses of the two Higgs candidate jets. Figure \ref{fig:MassPlotsDeltas} (left panel) shows the distribution of $\Delta_{h}$ for signal events and relevant backgrounds. Even after the overlap selections, the background distributions are significantly wider than the signal. Hence, in order to further suppress the background channels,  we impose a cut of 
\begin{equation}
|\Delta_{h}| < 0.1\,. 
\end{equation}

Upon identifying the $u$ and Higgs jets, we proceed with the reconstruction of the $U_h$ partner. The signal events are characterized by a distinct ``2 fat jet 2 light jet'' topology, a final state which represents somewhat of a combinatorial challenge (for each fat jet, two combinations with a light jet are possible).  In order to find the correct Higgs-light jet pairs, we construct four different combinations of invariant masses
\begin{equation}
	m^{U_h}_{ij}= \sqrt{(p^{h}_i + p^u_j)^2}\,, 
\end{equation}
where $p^{h}_i$ are the four momenta of the two $R=0.7$ jets which pass the Higgs tagging requirements and $p^u_j$ are the four momenta of the two hardest $r=0.4$ isolated from the Higgs jets by $\Delta R_{uh}  > 1.1.$ A correct Higgs-light jet pair then minimizes the value of
\begin{equation}
\Delta_{U_h} = \mathrm{min} \left[ |m^{U_h}_{11} -m^{U_h}_{22}|, |m^{U_h}_{12} - m^{U_h}_{21}|\right]\,.
\end{equation}

\begin{figure}[b]
\includegraphics[scale=0.37]{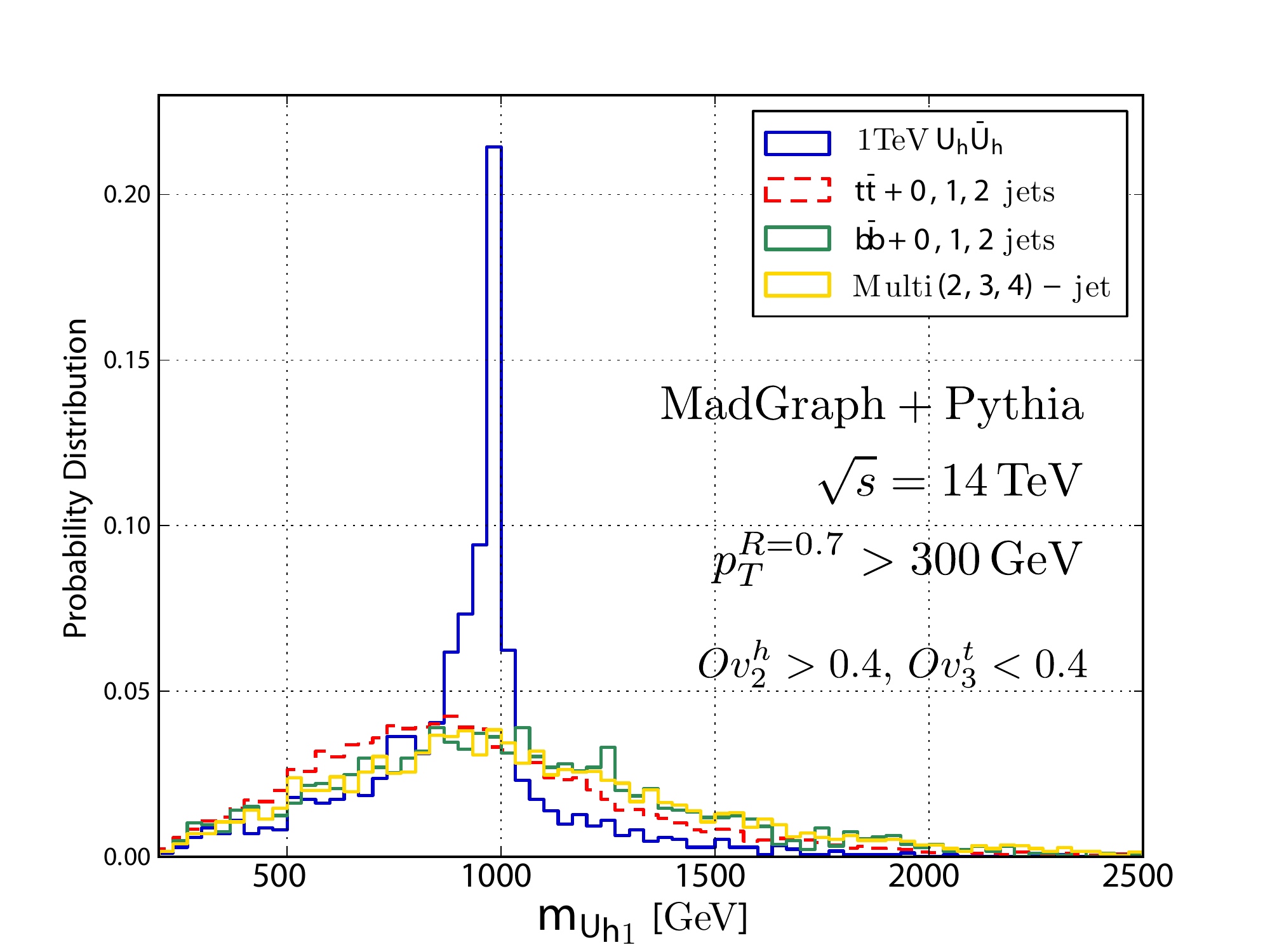}
\includegraphics[scale=0.37]{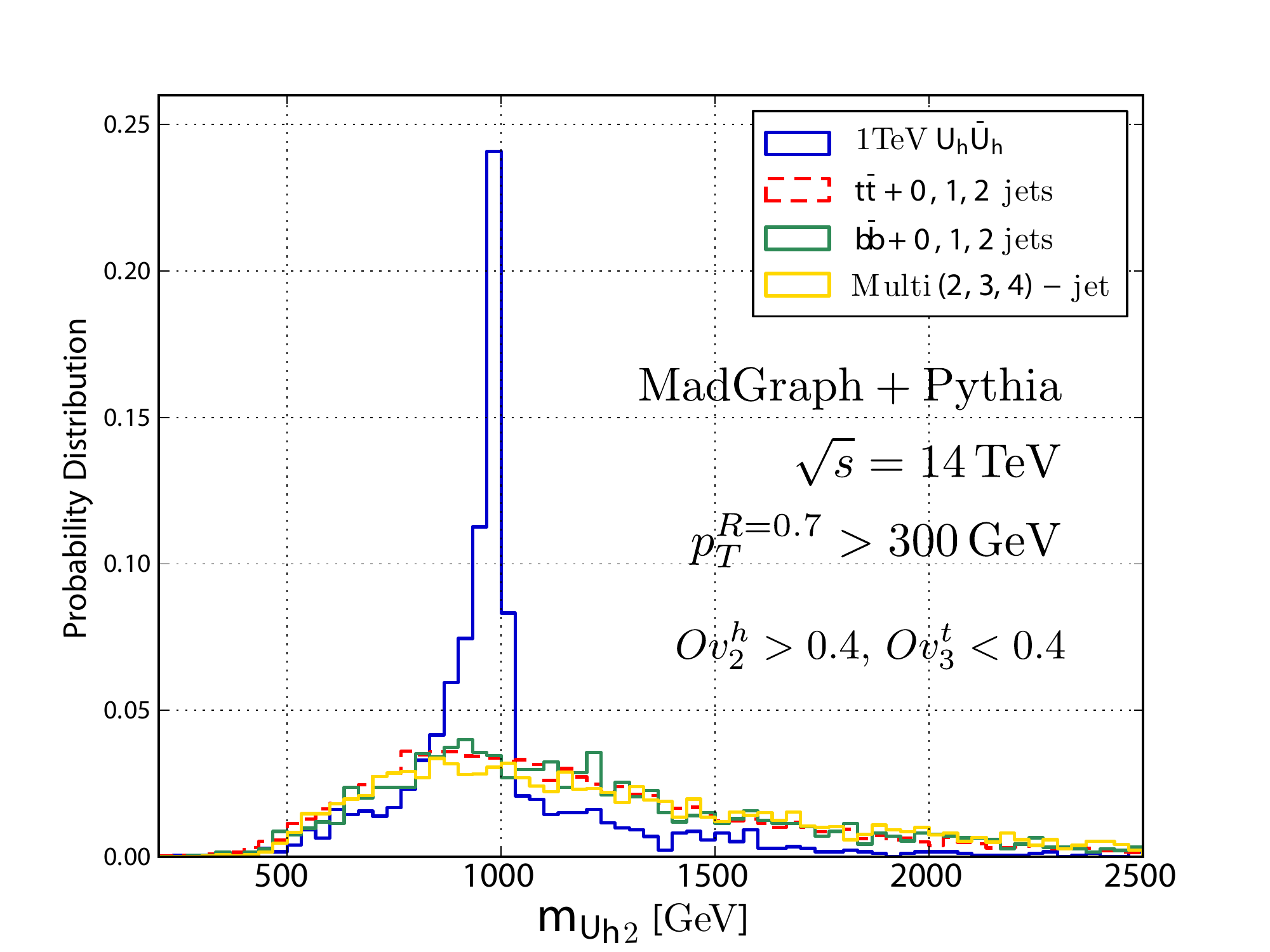}
\caption{Reconstructed mass of the $U_h$ partner for true mass $M_{U_h} = 1 \TeV$. Left panel shows the mass reconstruction from the hardest Higgs candidate jet and the light jet which minimized $\Delta_{U_h}$, while the right panel shows the corresponding distribution assuming the second hardest Higgs candidate.}
\label{fig:MassPlotsUh}
\end{figure}

Consequently, we take the configuration of Higgs - light jet  pair which minimizes $\Delta_{U_h}$ to construct $m_{U_{h1,2}}$, the masses of the two $U_h$ partners in the event. Figure \ref{fig:MassPlotsUh}  shows the reconstructed invariant mass distribution of the $U_h$ pair (assuming $M_{U_h} = 1 \TeV$) and the background distributions. The signal events show a prominent peak at the correct partner mass for both $U_h$ partners in the event, while the background distributions are smeared over a wide range of mass values. The results of Figure \ref{fig:MassPlotsUh} illustrate well the degree to which our proposal is able to resolve the mass of the $U_h$ partners. 

The value of $\Delta_{U_h}$ represents the minimum of a  mass asymmetry between the two reconstructed objets and hence utilizes the fact that the $U_h$ partners are pair produced. 
In addition to allowing us to overcome the combinatorial issues when reconstructing the $U_h$ partners, $\Delta_{U_h}$ provides another handle on the background channels. Because the $U_h$ partners are pair produced, we expect the value of $\Delta_{U_h}$ to peak at 0 for signal events, while we expect the background channels to be characterized by wider distributions of $\Delta_{U_h}$ as there is no kinematic feature in the background channels which would lead to a reconstruction of two same mass resonances. 
Figure \ref{fig:MassPlotsDeltas} (right panel) shows $\Delta_{U_h}$  distributions for signal and relevant backgrounds. As in the case of $\Delta_{h}$, the background distributions of $\Delta_{U_h}$ are significantly broader compared to the signal, hence providing another unique handle on the background channels. In order to exploit this feature,  we impose a cut on
\begin{equation}
	|\Delta_{U_h}| < 0.1\,,
\end{equation}
as a part of our event selection. 

Finally, since we are interested in $U_h$ partners with mass $O(1 \TeV)$, we require that both Higgs-light jet pairs pass the requirement
\begin{equation}
	m_{U_{h1,2}} > 800, 1000 \GeV,
\end{equation}
for the benchmark values of $M_{U_h} = 1, 1.2 \TeV$ respectively, where we construct the mass of $U_{h1}$ and $U_{h2}$ from Higgs-light jet pairs which minimize $\Delta U_h$. 

\subsection{LHC Run II Sensitivity to $U_h$ Partners of Mass $\sim 1\TeV$}
\label{sec:CutScheme}

In this section we investigate the ability of our cutflow proposal to detect $\sim 1 \TeV$ light quark partners which decay to a Higgs-light jet pairs  at the Run II of the LHC.
Table~\ref{tab:1TeVBC1}~and~\ref{tab:12TeVBC1} show the main results, with respect to the initial cross section values in Table~\ref{tab:HT1600}. For all results on significance we assume a nominal integrated luminosity of $35 \fb^{-1}$. 

\setlength{\tabcolsep}{4pt}
\vspace{.5cm}
\begin{table}[h]
\begin{center}
\begin{tabular}{|c|c|c|c|c|c|c|}
\hline
							&	$\sigma_{s}$ [fb]	&$\sigma_{t\bar{t}}$ [fb]	&$\sigma_{b\bar{b}}$ [fb]	&$\sigma_{\rm multi-jet}$ [fb] & $S/B$ 			& $S/ \sqrt{B}$		\\ \hline
	Preselection Cuts			&	6.8\phantom{0}	&	4.6 $ \times 10^{2}$	&	8.4 $ \times 10^{3}$	&	2.8 $ \times 10^{5}$	     & $2.4 \times 10^{-5}$  & 7.5 $ \times 10^{-2}$			\\ \hline
	Basic Cuts				&	1.2\phantom{0} &	4.6				&	16.0				&	6.8 $ \times 10^{2}$	     & 1.7 $ \times 10^{-3}$	& 2.7 $ \times 10^{-1}$			\\  \hline
	 $| \Delta_{h}  | < 0.1$  		 &     0.82 			&	1.7				&	6.5				&	2.8 $ \times 10^{2}$	     & 2.9	$ \times 10^{-3}$ & 2.9 $ \times 10^{-1}$		\\  \hline
	  $| \Delta_{U_h}  | < 0.1$ 		 &	0.56 			&	5.5 $\times 10^{-1}$	&	2.0				&	87.0				     & 6.3 $ \times 10^{-3}$	   & 3.5 $ \times 10^{-1}$		\\  \hline
	  $m_{U_{h1,2}} > 800$ GeV	&	0.50			&	3.6 $\times 10^{-1}$	&	1.6				&	67.0				     & 7.3 $ \times 10^{-3}$	   & 3.6 $ \times 10^{-1}$		\\  \hline	
	    $b$-tag  				&	0.34			&	4.4 $\times 10^{-2}$	&	1.1  $\times 10^{-2}$	&	1.5 $\times 10^{-2}$	     & \textbf{4.8}				& \textbf{7.5}			\\  \hline
\end{tabular}\par
\end{center}
\caption{ $M_{U_{h}} = 1$ TeV , $\sigma_s = 6.8 \fb$ , $\mathcal{L }= 35 \fb^{-1}$} \label{tab:1TeVBC1} 
\end{table}

\vspace{.5cm}
\begin{table}[h]
\begin{center}
\begin{tabular}{|c|c|c|c|c|c|c|}
\hline
						&$\sigma_{s}$ [fb]		  &$\sigma_{t\bar{t}}$ [fb]	&$\sigma_{b\bar{b}}$ [fb]	&$\sigma_{\rm multi-jet}$ [fb]	& $S/B$ 				& $S/ \sqrt{B}$		\\ \hline
Preselection Cuts			&	2.4\phantom{00}	  &	4.6 $ \times 10^{2}$	&	8.4 $ \times 10^{3}$	&	2.8 $ \times 10^{5}$		& $8.15 \times 10^{-6}$	& 2.6 $ \times 10^{-2}$ \\ \hline
	Basic Cuts 			& 	0.60\phantom{0}	  &	4.6				&	16.0				&	6.8 $ \times 10^{2}$		& 8.6 $ \times 10^{-4}$	& 1.4 $ \times 10^{-1}$	\\  \hline
	  $| \Delta_{h} | < 0.1$  	 &     0.39\phantom{0}	  &	1.7				&	6.5				&	2.8 $ \times 10^{2}$		& 1.4 $ \times 10^{-3}$	& 1.4 $ \times 10^{-1}$	\\  \hline
	  $| \Delta_{U_h} | < 0.1$ 	 &     0.27\phantom{0}	   &	5.5 $ \times 10^{-1}$	&	2.0				&	87.0					& 3.0 $ \times 10^{-3}$	& 1.7 $ \times 10^{-1}$	 \\  \hline
  $m_{U_{h1,2}} > 1$ TeV		&   0.22\phantom{0} 	   	&	1.9 $ \times 10^{-1}$	&	1.0				&	45.0					& 4.8	 $ \times 10^{-3}$	& 1.9 $ \times 10^{-1}$	\\  \hline	
	   $b$-tag  			 &  0.134				 &	2.2 $ \times 10^{-2}$	&	8.5 $ \times 10^{-3}$	&	1.2 $ \times 10^{-2}$		& \textbf{3.1}					& \textbf{3.8}			\\  \hline 
	\end{tabular}\par	
	\end{center}
\caption{ $M_{U_{h}} = 1.2$ TeV , $\sigma_s = 2.4 \fb$ , $\mathcal{L }= 35 \fb^{-1}$} \label{tab:12TeVBC1} 
\end{table}
\setlength{\tabcolsep}{6pt}

Our results show that boosted jet techniques combined with fat jet $b$-tagging and kinematic constraints of pair produced heavy particles can achieve $S/B > 1$ with signal significance of $\sim 7 \sigma$ at $35 \fb^{-1}$, assuming light quark partners of $ M_{U_h} =1 \TeV.$ The significance we obtain is sufficient to claim a discovery of $1 \TeV$ light quark partners. In addition, we find that probing masses higher than $1 \TeV$ will require more luminosity and will be challenging at Run II of the LHC. However,  even with $35 \fb^{-1}$ signal significance of more than $3\sigma$ is achievable for $M_{U_h} = 1.2 \TeV$, enough to rule out the model point. 

Requiring that there exist four fat jets with $p_T >300 \GeV$ in an event, together with our boosted Higgs tagging procedure result in an improvement of $S/B$ by roughly a factor of 70-100 at $\sim 20\%$ signal efficiency relative to the pre-selection cuts. Additional cuts on mass asymmetries improve $S/B$ by roughly of factor a 3 in total.

\begin{figure}[t]
\begin{tabular}{cc}
\includegraphics[scale=0.37]{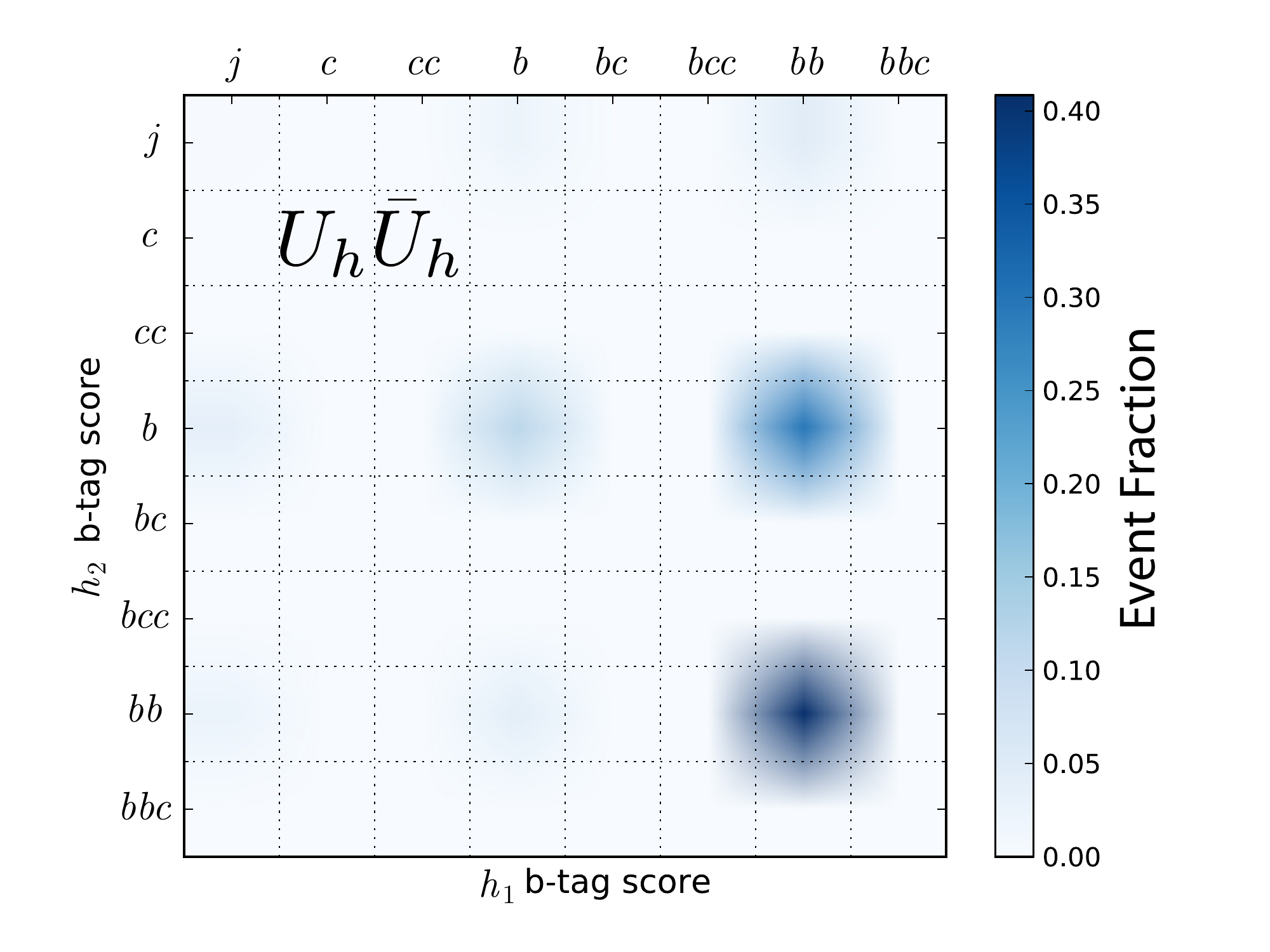}&
\includegraphics[scale=0.37]{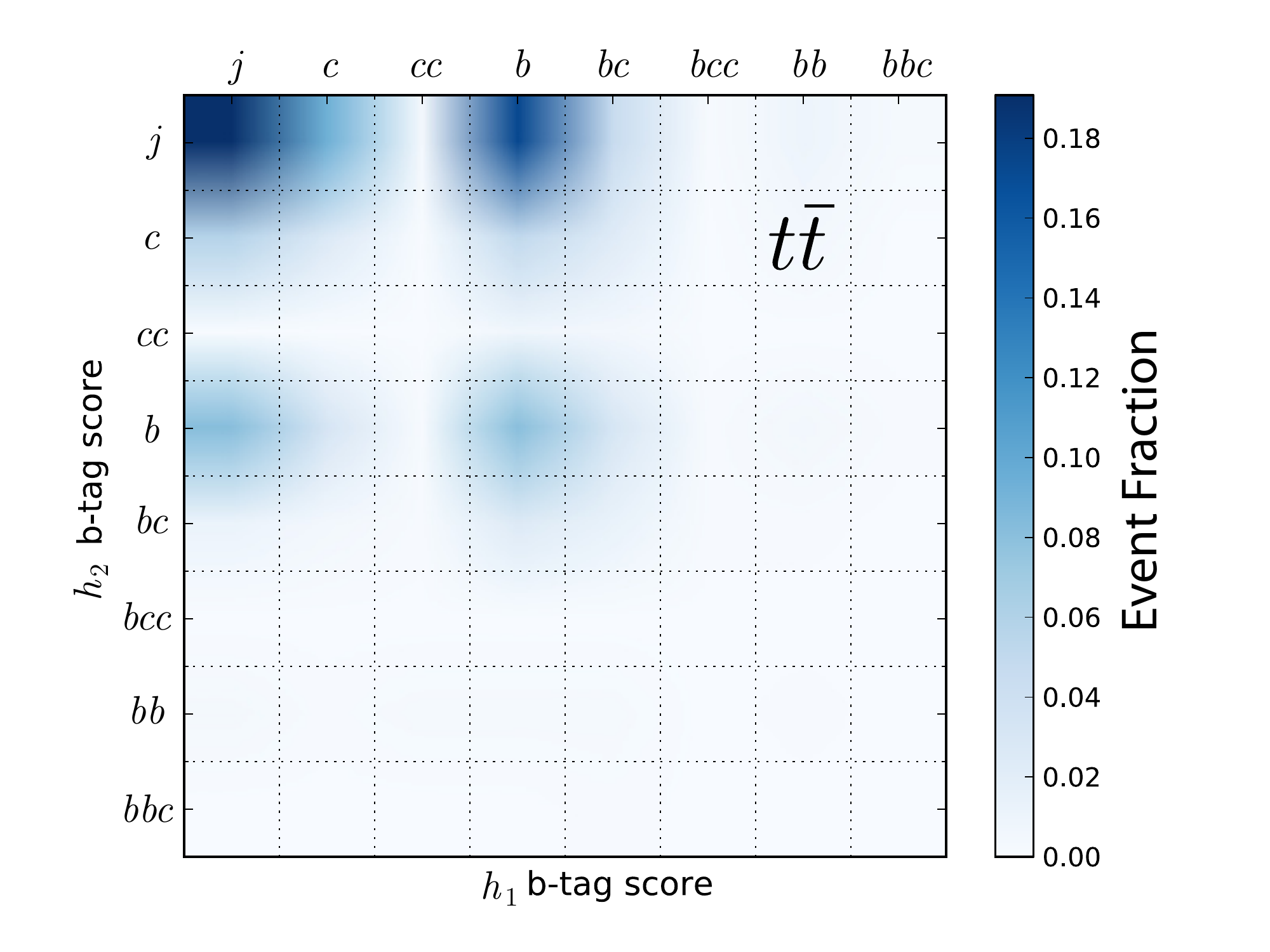}\\
\includegraphics[scale=0.37]{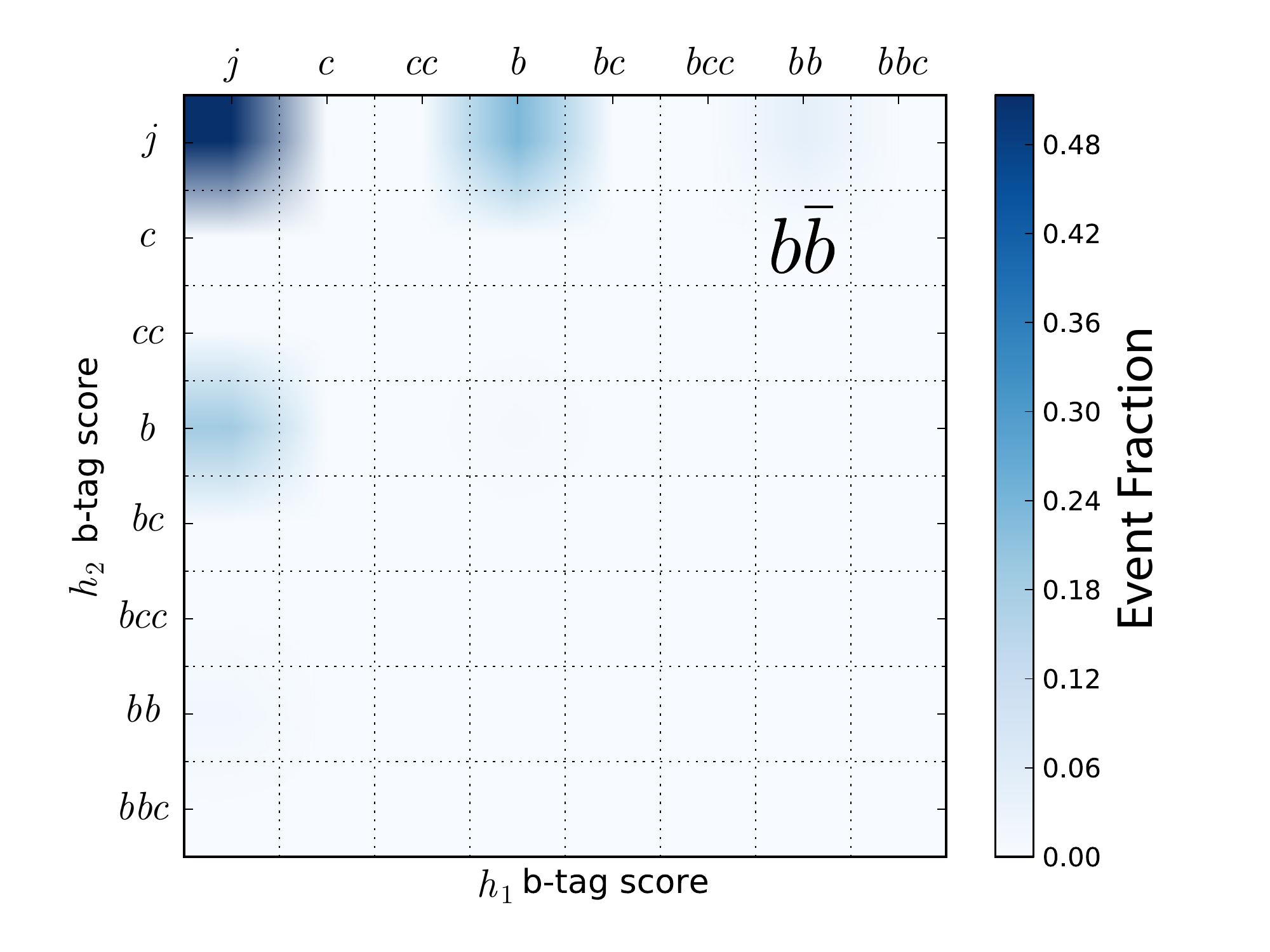}&
\includegraphics[scale=0.37]{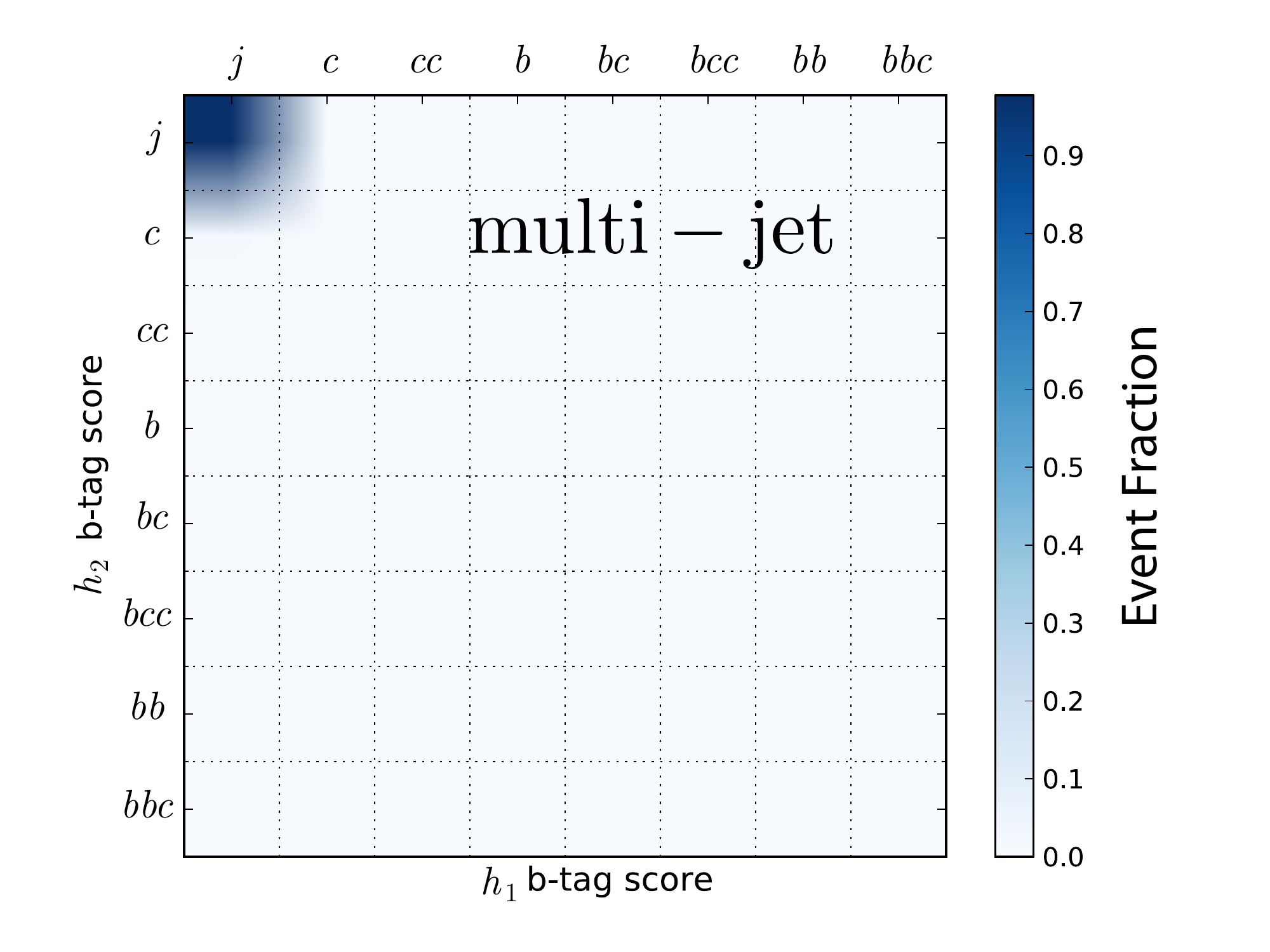}
\end{tabular}
\caption{$b$-tag score tables for the signal (top panel, left), `` $t \bar{t}$ +0,1,2 jets " (top panel, right), `` $b \bar{b}$ +0,1,2 jets" (bottom panel, left) and ``multi(2,3,4)-jet" (bottom panel, right), following the simplified $b$-tagging procedure of Section \ref{sec:b-tagging}. $h_{1,2}$ are the two highest $p_T$ fat jets which pass the Higgs tagging criteria of Section \ref{sec:TOM}. No $b$-tagging efficiencies have been applied to the results displayed in the plots.}
\label{fig:btagscores}
\end{figure}

The greatest improvement in both $S/B$ and $S/\sqrt{B}$ comes from fat jet $b$-tagging, where we find an enhancement by a factor of $\sim 500 -600 $ in $S/B$ and $15-20$ in signal significance. The improvement is largely due to the enormous suppression double fat-jet $b$-tagging exerts on the multi-jet and $b\bar{b}$ backgrounds, with the signal efficiency of $\sim 50\%$. The high rejection power of $b$-tagging can be understood well from results presented in Figure \ref{fig:btagscores}. The signal events almost always contain at least one $b$ quark in each of the fat jets which pass the boosted Higgs tagging criteria. Conversely, almost no multi-jet and $b\bar{b}$ events contain two ``Higgs like'' fat jets with each of the tagged heavy boosted objects containing a $b$-jet. The only background channel which seems to contain a significant fraction of events with both fat jets containing a proper $b$-tag is Standard Model $t\bar{t}$. Still, we find that only about $10\%$ of the $t\bar{t}$ events survive the double $b$-tagging criteria. 


\section{Conclusions}\label{sec:conclusions}
We studied the LHC Run II discovery potential for the light quark partners in composite Higgs models. As an example, we considered a simplified model based on the $SO(5)/SO(4)$ coset structure containing one up-type quark in the decoupling limit. Of particular interest were pair produced up-type quark partners of mass $\sim 1 \TeV$ which then decay into two boosted Higgses (which we take to decay further hadronically) and two hard jets -- a final state which can not be efficiently tagged and reconstructed by ``traditional'' jet techniques. We proposed a new event cut scheme, designed to exploit the characteristic features of the pair produced $U_h$ event topology. We found that a combination of $b$-tagging, jet substructure, and kinematic cuts resulting from the fact that quark partners are pair produced allows to suppress the large QCD backgrounds to a degree where $S/B > 1$ and $S/\sqrt{B}\sim 7$ is possible for quark partners of mass $1\TeV$ with $35 \fb^{-1}$ of integrated luminosity. Our results show that the LHC Run II could achieve sufficient sensitivity to light quark partners of mass $1 \TeV$ to claim discovery. Probing masses  higher than $1 \TeV$ using our proposed cut-scheme will be difficult at Run II of the LHC, yet with $35 \fb^{-1}$ we find that a signal significance of more than $3\sigma$ is achievable for $M_{U_h} = 1.2 \TeV$, sufficient to rule out the model point. 

The event selection procedure we propose begins by requiring the presence of four fat jets ($i.e. \, R=0.7$), two of which are tagged as Higgs candidates. We perform Higgs tagging by considering a combination of the Higgs two body peak overlap, $Ov_2^h$, and the top three body overlap $Ov_3^t$, where we require a \textit{lower} cut on $Ov_2^h$ and an \textit{upper} cut on $Ov_3^t$. The two-dimensional overlap analysis allows us to suppress the QCD backgrounds, including $t\bar{t}$, to a much better degree compared to the analysis utilizing only $Ov_2^h$. In addition to jet substructure tagging, we also require the two Higgs candidate jets to be $b$-tagged, as well as that the mass difference between the Higgs jets is small. Kinematics of heavy pair produced quark partners offer an additional handle on the background channels, and we require that the mass difference between the reconstructed $U_h$ partners also be small. 
The greatest improvement in the signal significance comes from $b$-tagging, as requiring two Higgs fat jets to be $b$-tagged diminishes the enormous multi-jet background. 

Our study represents a ``proof of principle'' that successful searches for TeV scale light quark partners decaying to $hj$ are possible at the Run II of the LHC. Further work is necessary to study the effects of pileup contamination on the results of the analysis. Yet, it is likely pileup effects will be manageable, even at $\sim 50$ interactions per bunch crossing. The TOM analysis of boosted jets is weakly susceptible to pileup at 50 interactions per bunch crossing \cite{Backovic:2013bga}, as long as the fat jet $p_T$ is corrected so that the appropriate template bin is used in the analysis. Alternatively, many issues with determining the jet $p_T$ in a high pileup environment could be bypassed by analyzing each jet with template sets at a range of transverse momenta. Effects of pileup on jet mass do not represent an issue for our event selection proposal, as the combination of $Ov_2^h$ and $Ov_3^t$ selections serves as an excellent intrinsic mass filter. Furthermore, recent experimental studies of Ref. \cite{Capeans:1291633} suggest that effects of pileup on $b$-tagging at LHC Run II will be under control. 

Future analyses using our event selection could also benefit from a detailed detector simulation.

\bigskip
\emph{Acknowledgements:} \\
The authors would like to thank the Weizmann theory group for the hospitality during the initial stages of this project.
SL acknowledges the Galileo Galilei Institute for Theoretical Physics in Florence, Italy, where part of this work was completed.
This work was supported by the National Research Foundation of Korea(NRF) grant funded by the Korea government(MEST) (No. 2012R1A2A2A01045722),
and Basic Science Research Program through the National Research Foundation of Korea(NRF) funded by the ministry of Education,
Science and Technology (No. 2013R1A1A1062597). This work is also supported by HTCaaS group of KISTI (Korea Institute of Science and Technology Information). SL and TF are also supported by Korea-ERC researcher visiting program through the National Research Foundation of Korea(NRF) (No. 2014K2a7B044399 and No. 2014K2a7A1044408).
JH is supported by the IBS Center for Theoretical Physics of the Universe.

\bigskip

\appendix
\section{Partially composite light quark partners with general $y_L,y_R$: Branching ratios of $U_h$
}\label{app:yLlyR}
In Section \ref{sec:model} we discussed a partially composite light quark partner model of a minimal composite Higgs model in which the elementary quarks as well as the composite partner quarks are embedded into a $\bf 5$ of $SO(5)$, and in which the $SO(4)$ singlet mass scale $M_1$ of one of the partners of the light quarks $u,d,s,c$ is lower than the remaining partners mass scales, such that the model can be described be the effective Lagrangian Eq.~(\ref{pcLag1}). In addition, we assumed dominance of the right-handed pre-Yukawa coupling of this quark partner, \ie $\,y_R \gg y_L$. In this case, the quark partner state decays dominantly into $h j$ and is described by the very simple effective Lagrangian Eq.~(\ref{Lpceff}) which we used for our further studies of the $hhjj$ signature at LHC run II. 

In the case of general $y_L,y_R$, the quark partner mass eigenstate has couplings to the $Z$, $W$, and Higgs bosons as given in Eqs.~(\ref{Zeff},\ref{Weff},\ref{lambdapceff}), which depend on the mixing angles $\varphi_{L,R}$ in the left- and right-handed sector. As the light SM quark mass is to be identified with $M_{u_l}$ given in Eq.~(\ref{Mul}), the product the mixing angles is tiny, the couplings in Eqs.~(\ref{Zeff},\ref{Weff},\ref{lambdapceff}) are small (unless an extreme hierarchy between $y_L$ and $y_R$ is chosen), and effect on $U_h$ production processes is negligible. However, changing the left- and right-handed mixing angles modifies the $U_h$ branching ratios.

The ``mixing'' couplings  Eqs.~(\ref{Zeff},\ref{Weff},\ref{lambdapceff}) imply decay channels of the $U_h$ into $Zu$, $Wd$, and $hu$ with partial decay widths \cite{Flacke:2013fya}
\bea
\Gamma_{U_h\rightarrow Z u}  =&M_{U_h}\frac{M^2_{U_h}}{m^2_Z}\frac{\left|g^L_{U_hu_hZ}\right|^2}{32 \pi}\Gamma_Z\,&= \frac{y^2_L}{2}\frac{M_{U_h}\Gamma_Z}{32\pi}+\mathcal{O}(m^2_{L,R}/M^2_1),\\
\Gamma_{U_h\rightarrow W d} =&M_{U_h}\frac{M^2_{U_h}}{m^2_W}\frac{\left|g^L_{U_hd_hW}\right|^2}{32 \pi}\Gamma_W\,&=  y^2_L \frac{M_{U_h}\Gamma_W}{32\pi}+\mathcal{O}(m^2_{L,R}/M^2_1),\\
\Gamma_{U_h\rightarrow h u}  =&M_{U_h}\frac{|\lambda_L|^2+|\lambda_R|^2}{64 \pi}\Gamma_h&= \left(\frac{y^2_L}{2} \cos^2(\epsilon)+y^2_R \sin^2(\epsilon)\right)\frac{M_{U_h}\Gamma_h}{32\pi}+\mathcal{O}(m^2_{L,R}/M^2_1)\,,\nonumber\\
\eea
where $\Gamma_{W,Z,h}= 1+\mathcal{O}(\frac{m^2_{W/Z/h}}{M^2_{U_h}})$ are kinematic functions, and we used the expressions for the couplings Eqs.~(\ref{Zeff},\ref{Weff},\ref{lambdapceff}), mixing angles Eq.~(\ref{eqmixangles}), as well as $246 \GeV \equiv v = f \sin(\langle h \rangle / f) \equiv  f \sin(\epsilon)$.

Thus, the Higgs decay channel dominates in the limit $y_R\gg y_L$, where $U_h$ decays through the right-handed channel, where  while for $y_L \cos(\epsilon) \gg \sqrt{2} y_R \sin(\epsilon)$ decays through the left-handed channel dominate, which leads to branching ratios   $\Gamma_{U_h\rightarrow W d}: \Gamma_{U_h\rightarrow Z u}: \Gamma_{U_h\rightarrow h u}$ of $\sim 2: \sim 1: \sim 1$.
In the latter parameter regime, the discovery and exclusion reach of the model purely the $hhjj$ channel (as discussed in this article) is substantially reduced because the cross-section of this channel is reduced by a factor of $\sim 16$. However, decays of the $U_h$ into $Wd$ and $Zu$ imply a variety of final states ($WWjj$, $ZZjj$, $WjZj$, $Wjhj$, ect.  with hadronic or leptonic $W$ and $Z$ decays) in which the model can be tested. A combination of such searches can be expected to lead to sensitivity to higher masses at LHC run II for studies of these signatures in composite Higgs or other models.

\bibliographystyle{apsrev}
\bibliography{draft_paper}
\end{document}